\documentclass[aps,prd,superscriptaddress,amsmath,twocolumn,10pt]{revtex4-1}
\usepackage{color,hangcaption,hhline,psfrag,rotating,amssymb}
\usepackage[hang,nooneline]{subfigure}
\usepackage{dcolumn}
\usepackage{verbatim}

\newcommand{\nn}{\nonumber}
\newcommand{\be}{\begin{equation}}
\newcommand{\ee}{\end{equation}}
\newcommand{\bea}{\begin{eqnarray}}
\newcommand{\eea}{\end{eqnarray}}

\newcommand{\stat}{ {}_{\mbox{\tiny stat}}}
\newcommand{\syst}{ {}_{\mbox{\tiny syst}}}

\begin{document}
\title{Precise heavy-light meson masses and hyperfine splittings from lattice QCD including charm quarks in the sea}

\author{R.~J.~Dowdall}
\email[]{Rachel.Dowdall@Glasgow.ac.uk}
\author{C.~T.~H.~Davies}
\email[]{c.davies@physics.gla.ac.uk}
\affiliation{SUPA, School of Physics and Astronomy, University of Glasgow, Glasgow, G12 8QQ, UK}
\author{T.~C.~Hammant}
\author{R.~R.~Horgan}
\affiliation{DAMTP, University of Cambridge, Wilberforce Road, Cambridge CB3 0WA, UK}
\collaboration{HPQCD collaboration}
\homepage{http://www.physics.gla.ac.uk/HPQCD}

\date{\today}

\begin{abstract}
We present improved results for the $B$ and $D$ meson spectrum from lattice QCD including the effect of $u/d,s$ and $c$ quarks in the sea. For the $B$ mesons the Highly Improved Staggered Quark action is used for the sea and light valence quarks and NonRelativistic QCD for the $b$ quark including $\mathcal{O}(\alpha_s)$ radiative corrections to many of the Wilson coefficients for the first time.
The $D$ mesons use the Highly Improved Staggered Quark action for both valence quarks on the same sea.
We find $M_{B_s} - M_{B}=84(2)$ MeV, $M_{B_s}=5.366(8)$ GeV, $M_{B_c}=6.278(9)$ GeV, $M_{D_s}=1.9697(33)$ GeV, and $M_{D_s}-M_{D}=101(3)$ MeV.
Our results for the $B$ meson hyperfine splittings are $M_{B^*}-M_{B}=50(3)$ MeV, $M_{B_s^*}-M_{B_s}=52(3)$ MeV, in good agreement with existing experimental results. 
This demonstrates that our perturbative improvement of the NRQCD chromo-magnetic coupling works for both heavyonium and heavy-light mesons. We predict $M_{B_c^*}-M_{B_c}=54(3)$ MeV. 
We also present first results for the radially excited $B_c$ states as well as the orbitally excited scalar 
$B_{c0}^*$ and axial vector $B_{c1}$ mesons.
\end{abstract}

%\pacs{}
%\keywords{}

\maketitle

%%%%%%%%%%%%%%%%%%%%%%%%%%%%%%%%%%%%%%%%%%%%%%%%%%%%%%%%%%%%%%%%%
%
\section{Introduction}
\label{sec:intro}
%
%%%%%%%%%%%%%%%%%%%%%%%%%%%%%%%%%%%%%%%%%%%%%%%%%%%%%%%%%%%%%%%%%+

Lattice QCD calculations have become an essential part of $B$ physics phenomenology~\cite{Davies:2012qf}, providing increasingly precise determinations of decay constants and mixing parameters needed, along with experiment, in the determination of CKM matrix elements. Since these calculations can now give stringent constraints on the CKM unitarity triangle, currently resulting in tension at a few sigma level \cite{Laiho:2012ss}, it is important to check that all systematic errors have been correctly accounted for. With this in mind we present a new study of the B-meson spectrum that provides a good check of recent improvements that have been made in our discretization of the QCD Lagrangian.
The possibility of more $B$ states being found at experiments such as LHCb also gives us the opportunity for further tests of QCD in the nonperturbative regime.
We emphasise that all parameters for this calculation, including quark 
masses and the lattice spacing, have already been determined elsewhere~\cite{Dowdall:2011wh} making this a parameter free test of lattice QCD.

This test is made possible by the use of NonRelativistic QCD (NRQCD) for the $b$ quark, which has the advantage that the same action can be used for both bottomonium and $B$-meson calculations. HPQCD recently computed the one loop radiative corrections to many of the coefficients in the NRQCD action~\cite{Hammant:2011bt, Dowdall:2011wh} and studied the effect of these improvements on the bottomonium spectrum in \cite{Dowdall:2011wh}. Systematic errors were significantly reduced in a number of quantities, including the hyperfine splitting, and the first QCD prediction of the D-wave spin splittings was made 
\cite{Daldrop:2011aa}. This analysis used new gluon configurations~\cite{Bazavov:2010ru} generated by the MILC collaboration with 2+1+1 flavours of HPQCD's Highly Improved Staggered Quarks (HISQ)\cite{Follana:2006rc} in the sea and including $n_f\alpha_s a^2$ improvements 
to the gluon action~\cite{Hart:2008sq}. We use the same gluon configurations here. 

For the $u$, $d$, $s$ and $c$ valence quarks in our calculation 
we use the same HISQ action as for the sea quarks. 
The advantage of using HISQ is that $am_q$ discretisation errors are under 
sufficient control that it can be used both for light 
and for $c$ quarks~\cite{Follana:2006rc}. 
Both the NRQCD and the HISQ actions are also numerically very cheap which 
means we are able to perform a very high statistics calculation 
to combat the signal to noise ratio problems that arise in simulating B-mesons. 
The same $u/d,s,c$ HISQ quark propagators used in the $B$ mesons can also be used to calculate the masses of pseudoscalar charmed mesons which we also present here. Our results are precise enough that it is possible to distinguish the heavy quark dependence of splittings such as the $M_{D_s}-M_D$ and $M_{B_s}-M_B$.

We begin by outlining the methods used in our lattice calculation, which are similar to \cite{Dowdall:2011wh,Gregory:2010gm}.
The $B_s,B_c$ and $B$ meson masses and the radially excited $B_c^{'}$ are presented in Sec.~\ref{sec:mesons}, hyperfine results are given in Sec.~\ref{sec:hyperfines}, axial vector and scalar $B$-mesons are discussed in \ref{sec:axialmesons}.
Sec.~\ref{sec:discussion} compares our results to earlier NRQCD-HISQ ones on $n_f=2+1$ 
configurations including asqtad sea quarks~\cite{Gregory:2010gm} and to 
calculations using the HISQ action for $b$ quarks~\cite{McNeile:2011ng, McNeile:2012qf}. 
Sec.~\ref{sec:conclusions} gives our 
conclusions, including an updated spectrum for gold-plated mesons from lattice QCD.

%%%%%%%%%%%%%%%%%%%%%%%%%%%%%%%%%%%%%%%%%%%%%%%%%%%%%%%%%%%%%%%%
%
\section{Lattice calculation}
\label{sec:lattice}
%
%%%%%%%%%%%%%%%%%%%%%%%%%%%%%%%%%%%%%%%%%%%%%%%%%%%%%%%%%%%%%%%%%

Our calculation uses five ensembles of gluon configurations generated by the MILC collaboration \cite{Bazavov:2010ru}. 
These are $n_f = 2+1+1$ configurations that include the effect of light, strange and charm quarks with the HISQ action and a Symanzik improved gluon action with coefficients correct through $\mathcal{O}(\alpha_s a^2,n_f\alpha_s a^2)$ \cite{Hart:2008sq}. The lattice spacing 
values range from $a=0.15$fm to $a=0.09$fm. The configurations have accurately tuned 
sea strange quark masses and sea light quark masses ($m_u=m_d=m_l$) with ratios to the strange 
mass of $m_l/m_s=0.1$ and $0.2$, which correspond to pions of mass 220-315 MeV. Having sea quark masses close to the physical point is particularly important for studies of the $B$ meson where chiral extrapolations make up a substantial portion of the final error.

In Ref.~\cite{Dowdall:2011wh} we accurately determined the lattice spacings using the $\Upsilon(2S-1S)$ splitting and the decay constant of 
the fictitious $\eta_s$ particle, a pseudo-scalar $s\bar s$ meson whose valence 
quarks are not allowed to annihilate on the lattice~\cite{Davies:2009tsa}. 
Agreement was shown between these methods in the continuum limit. In this paper we use the $\Upsilon(2S-1S)$ lattice spacings. The details of each ensemble, including the sea quark masses and spatial volumes, are given in Table \ref{tab:params}. All ensembles were fixed to Coulomb gauge.

\begin{table*}
\caption{
Details of the five gauge ensembles used in this calculation \cite{Bazavov:2010ru}. $\beta$ is the gauge coupling, $a_{\Upsilon}$ is the lattice spacing as determined by the $\Upsilon(2S-1S)$ splitting in \cite{Dowdall:2011wh}, where the three errors are statistics, NRQCD systematics and experiment. 
$am_l,am_s$ and $am_c$ are the sea quark masses, $L/a \times T/a$ gives the spatial and temporal extent of the lattices and $n_{{\rm cfg}}$ is the number of configurations in each ensemble. 
The ensembles 1 and 2 will be referred to in the text as ``very coarse'', 3 and 4 as ``coarse'' and 5 as ``fine''. 
}
\label{tab:params}
\begin{ruledtabular}
\begin{tabular}{lllllllll}
Set & $\beta$ & $a_{\Upsilon}$ (fm) 	& $am_{l}$ & $am_{s}$ & $am_c$ & $L/a \times T/a$ & $n_{{\rm cfg}}$  \\
\hline
1 & 5.80 & 0.1474(5)(14)(2)  & 0.013   & 0.065  & 0.838 & 16$\times$48 & 1020 \\
2 & 5.80 & 0.1463(3)(14)(2)  & 0.0064  & 0.064  & 0.828 & 24$\times$48 & 1000 \\
\hline
3 & 6.00 & 0.1219(2)(9)(2)   & 0.0102  & 0.0509 & 0.635 & 24$\times$64 & 1052 \\
4 & 6.00 & 0.1195(3)(9)(2)   & 0.00507 & 0.0507 & 0.628 & 32$\times$64 & 1000 \\
\hline
5 & 6.30 & 0.0884(3)(5)(1)   & 0.0074  & 0.037  & 0.440 & 32$\times$96 & 1008 \\
\end{tabular}
\end{ruledtabular}
\end{table*}

Light, strange and charm quark propagators were generated using the HISQ action, the masses used are given in Table \ref{tab:hisqparams}. 
In Ref.~\cite{Dowdall:2011wh} accurate strange quark masses were 
given for each ensemble, tuned from the mass of the $\eta_s$ meson, 
which was determined from $K$ and $\pi$ meson masses to be 0.6893(12) GeV. 
The values of $am_s^{\mathrm{val}}$ in Table \ref{tab:hisqparams} correspond to these. 
Mistuning of the strange quark mass was a major source of error in Ref.~\cite{Gregory:2010gm} which will not be present in this calculation.
The light valence quarks are taken to have the same masses as in the sea. 

Charm quark masses are tuned by matching the mass of the $\eta_c$ to experiment. The experimental value is shifted by 2.6 MeV for missing electromagnetic effects and 2.4 MeV for not allowing it to annihilate to gluons, giving 2.985(3) GeV~\cite{Davies:2009tsa}. 
The $\epsilon_{\mbox{\tiny Naik}}$ term in the action is not negligible for charm quarks and we use the tree level formula given in \cite{newfds}, the values appropriate to our masses are given in Table \ref{tab:hisqparams}.

\begin{table}
\caption{ 
The parameters used in the generation of the HISQ propagators. $am_q^{\rm val}$ are the valence quark masses and $\epsilon_{\mbox{\tiny Naik}}$ is the coefficient of the Naik term in the action. On set 4 $\epsilon_{\mbox{\tiny Naik}}$ is very slightly wrong - it should be -0.224.  
\label{tab:hisqparams}
}
\begin{ruledtabular}
\begin{tabular}{lllll}
Set & $am_l^{\rm val}$ & $am_s^{\rm val}$ & $am_c^{\rm val}$ & $\epsilon_{\mbox{\tiny Naik}}$ \\
\hline
1 & 0.013   & 0.0641 & 0.826  & -0.345 \\
2 & 0.0064  & 0.0636 & 0.818  & -0.340\\
\hline
3 & 0.01044 & 0.0522 & 0.645  & -0.235  \\ 
4 & 0.00507 & 0.0505 & 0.627  & -0.222  \\ 
\hline
5 & 0.0074  & 0.0364 & 0.434  & -0.117  \\ 
\end{tabular}
\end{ruledtabular}
\end{table}

%%%%%%%%%%%%%%%%%%%%%%%%%%%%%%%%%%%%%%%%%%%%%%%%%%%%%%%%%%%%%%%%%

The velocity of a $b$ quark in a bound state is typically very small; $v^2=0.1$ in bottomonium and $v^2$ varies from 0.01 to 0.04 in heavy light systems containing a $b$ quark. This makes NRQCD \cite{Thacker:1990bm} a suitable effective field theory for handling $b$ quarks. It also has a number of other advantages. By construction, we are able to perform calculations at relatively coarse lattice spacings since discretisation errors are not set by powers of the quark mass as in a relativistic theory. Generation of propagators is very fast since in NRQCD they can simply be generated by time evolution with a given Hamiltonian. 
The other major benefit is that NRQCD can be used for both heavy-heavy and heavy-light mesons. All free parameters in this calculation were previously tuned using the statistically more precise bottomonium spectrum in \cite{Dowdall:2011wh}, meaning that all results here are parameter free tests of QCD.

These advantages come at a price. NRQCD is non-renormalisable 
because operators of dimension greater than four are included 
in the action, rather than being evaluated as operator 
insertions as in HQET. 
This means that the continuum limit $a \rightarrow 0$ cannot be 
taken. 
This does not mean, however, that physical results cannot be extracted.
Because NRQCD is an effective theory, continuum results can be inferred 
from fits to calculations in its regime of validity, where $am_b >1$.
We discuss this in Sec. \ref{subsubsec:Bsmeson-fits}. 
As finer lattices become more readily available on which $am_b < 1$, 
other methods~\cite{McNeile:2011ng} may become more appropriate than NRQCD. 
In the meantime, however, NRQCD remains the easiest and best 
way to access the full  range of heavy quark physics in lattice QCD. 

The NRQCD Hamiltonian we use is given by \cite{Lepage:1992tx}:
 \begin{eqnarray}
 aH &=& aH_0 + a\delta H; \nonumber \\
 aH_0 &=& - \frac{\Delta^{(2)}}{2 am_b}, \nonumber \\
a\delta H
&=& - c_1 \frac{(\Delta^{(2)})^2}{8( am_b)^3}
            + c_2 \frac{i}{8(am_b)^2}\left(\bf{\nabla}\cdot\tilde{\bf{E}}\right. -
\left.\tilde{\bf{E}}\cdot\bf{\nabla}\right) \nonumber \\
& & - c_3 \frac{1}{8(am_b)^2} \bf{\sigma}\cdot\left(\tilde{\bf{\nabla}}\times\tilde{\bf{E}}\right. -
\left.\tilde{\bf{E}}\times\tilde{\bf{\nabla}}\right) \nonumber \\
 & & - c_4 \frac{1}{2 am_b}\,{\bf{\sigma}}\cdot\tilde{\bf{B}}  
  + c_5 \frac{\Delta^{(4)}}{24 am_b} \nonumber \\
 & & -  c_6 \frac{(\Delta^{(2)})^2}{16n(am_b)^2} .
\label{eq:Hamiltonian}
\end{eqnarray}
Here $\nabla$ is the symmetric lattice derivative and $\Delta^{(2)}$ and 
$\Delta^{(4)}$ the lattice discretization of the continuum $\sum_iD_i^2$ and 
$\sum_iD_i^4$ respectively. $am_b$ is the bare $b$ quark mass. 
$\bf \tilde{E}$ and $\bf \tilde{B}$ are the chromoelectric 
and chromomagnetic fields calculated from an improved clover term~\cite{Gray:2005ur}.
The $\bf \tilde{B}$ and $\bf \tilde{E}$ are made anti-hermitian 
but not explicitly traceless, to match the perturbative calculations 
done using this action.  

The coefficients $c_i$ in the action are unity at tree level but radiative corrections cause them to depend on $am_b$ at higher orders in 
$\alpha_s$. These were calculated for the relevant $b$ quark masses using lattice perturbation theory in \cite{Dowdall:2011wh} and the values used in this paper are given in Table \ref{tab:wilsonparams}. 
A major improvement in this work is the inclusion of one loop radiative corrections to $c_4$ \cite{Hammant:2011bt} which controls the hyperfine splitting between the vector and pseudo-scalar states. We show in Sec. \ref{sec:hyperfines} that this leads to accurate results for $b$-light hyperfine splittings 
in keeping with the results of~\cite{Dowdall:2011wh} for bottomonium.

The tuning of the $b$ quark mass on these ensembles was 
discussed in \cite{Dowdall:2011wh}. We use the spin-averaged kinetic 
mass of the $\Upsilon$ and $\eta_b$ and take the experimental 
value to which we tune to be 9.445(2) GeV. This allows for electromagnetism 
and $\eta_b$ annihilation effects missing from our calculation~\cite{Gregory:2010gm}.
Note that we no longer have to apply a shift for missing charm quarks 
in the sea~\cite{Gregory:2010gm}. 
The values used in this calculation are tuned on that basis 
and given in Table \ref{tab:upsparams} along with other parameters.

\begin{table}
\caption{ The coefficients $c_1$, $c_5$, $c_4$ and $c_6$ used in the NRQCD action \ref{eq:Hamiltonian}. $c_2$ and $c_3$ are set to 1.0.
}
\label{tab:wilsonparams}
\begin{ruledtabular}
\begin{tabular}{lllll}
Set & $c_1$ & $c_5$ & $c_4$ & $c_6$ \\
\hline
very coarse 	& 1.36 & 1.21 & 1.22 & 1.36 \\
coarse 		& 1.31 & 1.16 & 1.20 & 1.31 \\
fine 		& 1.21 & 1.12 & 1.16 & 1.21 \\
\end{tabular}
\end{ruledtabular}
\end{table}

The calculation of NRQCD-HISQ two point functions with 
stochastic noise sources uses the method developed 
in~\cite{Gregory:2010gm} to allow spin-information to be 
added into the HISQ propagators so that the correct 
$J^{PC}$ NRQCD-light correlators can be made. 
Once HISQ propagators have been made 
with a source time-slice of random numbers we can no longer apply 
the `staggering matrix', $\Omega(x) = \prod_{\mu=1}^4 \gamma_{\mu}^{x_{\mu}}$, 
at the source to convert them to naive 
quark propagators with spin as would be 
used in the 
original method for combining staggered 
and non-staggered quarks~\cite{Wingate:2002fh}. 
Instead we include the staggering 
matrix  at the source of the NRQCD propagators along with the same 
time-slice of random numbers~\cite{Gregory:2010gm, Na:2012kp}.
 
We also use exponentially smeared quark sources, which take 
form $\exp(-r/a_{sm})$ as a function of radial distance, for the NRQCD 
propagators. These use two different radial sizes, $a_{sm}$, on each 
ensemble as given in Table~\ref{tab:upsparams}.
Correlators were calculated at 16 time sources on each configuration and the calculation was repeated with the heavy quark propagating in the opposite time direction. All correlators on the same ensemble were binned to avoid underestimating the errors.
Our method also requires the calculation of $\Upsilon$ and $\eta_b$ correlators to subtract the unphysical ground state energy of NRQCD, for details see \cite{Dowdall:2011wh}.

\begin{table}
\caption{ Parameters used in the NRQCD action.
$am_b$ is the 
bare $b$ quark mass and $u_{0L}$ the Landau link tadpole-improvement 
factor used in the NRQCD action \cite{Lepage:1992xa}.  $n_{{\rm cfg}}$ gives the number 
of configurations used in each ensemble. 16 time sources were used on each 
configuration. 
The column $a_{sm}$ gives the size parameters of the quark 
smearing functions, which take the form $\exp(-r/a_{sm})$. 
$a_{sm}$ kept approximately constant in physical units.
}
\label{tab:upsparams}
\begin{ruledtabular}
\begin{tabular}{lllll}
Set & $am_b$ & $u_{0L}$ & $n_{t}$  & $a_{sm}$  \\
\hline
1 & 3.297 & 0.8195  & 16  & 2.0,4.0  \\
2 & 3.263 & 0.82015 & 16  &  \\
\hline
3 & 2.66 & 0.834  & 16  & 2.5,5.0 \\ 
4 & 2.62 & 0.8349 & 16  &  \\ 
\hline
5 & 1.91 & 0.8525 & 16  & 3.425,6.85  \\ 
\end{tabular}
\end{ruledtabular}
\end{table}

$B$ meson energies are extracted from the two-point functions using a simultaneous multi-exponential Bayesian fit \cite{gplbayes,fitcode} to the form
\begin{eqnarray}
\label{eq:fitform}
C_{\rm meson}(i,j,t_0;t)&=& \sum^{N_{\exp}}_{k=1} b_{i,k}b^*_{j,k}e^{-E_{k}(t-t_0)
}\\
 &-& \sum^{N_{\exp}-1}_{k^\prime=1}   d_{i,k^\prime}d^*_{j,k^\prime}(-1)^{(t-t_0)
}e
^{-E^\prime_{k^\prime}(t-t_0)}.\nonumber
\end{eqnarray}
The priors on the energy splittings $E_{n+1}-E_n$ are 600(300) MeV and the priors on the ground states are estimated from previous results with a width of 300 MeV. The priors on the amplitudes are 0.1(1.0)
and the fit includes points from some $t_0$ to $L_t/2$, half the temporal extent of the lattice. 
$i$ and $j$ label the different source and sink smearing functions used in the correlator.
$t_0$ is taken from $7-8$ on the fine ensemble, $6-8$ on coarse and for very coarse the $B$ and $B_s$ are fit from $t_0=4-8$ but the $B_c$ fits started at $t_0=14$ in order to obtain an acceptable fit. The $B,B_s$ and $B_c$ are fit separately but all vector and pseudo-scalar correlators for each meson are included in the same fit. Scalar and axial vector states are obtained from the oscillating terms 
(i.e. the $E^\prime_{k^\prime}$) in Eq.~(\ref{eq:fitform}). 
The oscillating terms 
correspond to opposite parity states made by the time-doubled quark 
and are typically present 
in meson correlators made from staggered quarks. 
 
%%%%%%%%%%%%%%%%%%%%%%%%%%%%%%%%%%%%%%%%%%%%%%%%%%%%%%%%%%%%%%%%%
%
\section{Meson masses}
\label{sec:mesons}
%
%%%%%%%%%%%%%%%%%%%%%%%%%%%%%%%%%%%%%%%%%%%%%%%%%%%%%%%%%%%%%%%%%

We begin with results for pseudo-scalar mesons. Hyperfine splittings are 
discussed in Sec.~\ref{sec:hyperfines} and scalars and axial vectors 
in Sec.~\ref{sec:axialmesons}.

%%%%%%%%%%%%%%%%%%%%%%%%%%%%%%%%%%%%%%%%%%%%%%%%%%%%%%%%%%%%%%%%%
%
\subsection{The $B_s$ meson}
\label{subsec:Bsmeson}
%
%%%%%%%%%%%%%%%%%%%%%%%%%%%%%%%%%%%%%%%%%%%%%%%%%%%%%%%%%%%%%%%%%
In NRQCD meson energies have an unphysical energy shift and we must consider energy splittings in order to compare with experiment. We subtract half the spin average $aE_{b\bar b}$ of the $\Upsilon$ and $\eta_b$ ground state energies from $aE_{B_s}$
\begin{equation}
\label{EXPI}
\Delta_{B_s}= \left(aE_{B_s} - \frac{1}{2}aE_{b\overline{b}}\right)_{\rm latt}a^{-1}. 
\end{equation}
From this we can reconstruct $M_{B_s}$ using
\begin{equation}
M_{B_s,{\rm latt}} = \Delta_{B_s} + \frac{1}{2}M_{b\overline{b}, {\rm phys}}
\label{eq:mbsrecon}
\end{equation}
where $M_{b\overline{b}, {\rm phys}}=9.445(2)$ is the relevant experimental value. 

Our results for $aE_{B_s} $ and  $aE_{b\bar b}$ are 
given in Table \ref{tab:bsresults}.  
Our $b$ and $s$ quark masses are well-tuned here. 
Nevertheless we allow small adjustments to $\Delta_{B_s}$ to 
allow for mistuning. These are based on previous determinations 
of the linear slope of $\Delta_{B_s}$ with appropriate meson 
mass, $M_{b\bar b}$ for $b$ and $M_{\eta_s}^2$ for $s$.  
In \cite{Gregory:2010gm}, the slope of $\Delta_{B_s}$ against $M_{b\bar b}$ 
was found to be 0.017 using two values of $am_b$ on a very 
coarse ensemble. By comparing our spin averaged kinetic masses to 
the experimental value on each ensemble, we obtain 
the shift $\Delta_{M_{b\bar b}}$ that needs to be applied to  $\Delta_{B_s}$ to give the value at the correct $b$ quark mass. 
Using two values of $am_s$ on set 1, we find that the 
slope of $\Delta_{B_s}$ with $M_{\eta_s}^2$ is 0.24(4), 
consistent with previous results~\cite{Gregory:2010gm, McNeile:2011ng}. 
Comparing $M_{\eta_s}$ on each ensemble to the physical value 
of 0.6893(12) GeV in \cite{Dowdall:2011wh}  gives the tuning 
shift  $\Delta_{M_{\eta_s}^2}$. This is significantly smaller 
in all cases than the lattice spacing error in $\Delta_{B_s}$. 
The error on both shifts is taken to be half the shift itself.

The splittings $\Delta_{B_s}$ before shifts are applied 
are listed in Table \ref{tab:Bstune} 
along with the shifts due to mistuning.
$M_{B_s,{\rm latt}}$ is plotted in Fig. \ref{fig:Bs}. 
The error is dominated by that from the lattice spacing uncertainty. 
This error would be reduced if we constructed an energy 
difference which was much smaller, for example subtracting 
$M_{\eta_s}$ from both sides of Eq.~(\ref{EXPI}). However the 
resulting quantity would then be very sensitive to the $s$ quark 
mass, so we do not do this here. 
As Fig.~\ref{fig:Bs} shows, no significant 
lattice spacing or sea quark mass dependence 
is visible in our results for $M_{B_s,{\rm latt}}$.

\begin{table*}
\begin{ruledtabular}
\begin{tabular}{lllllllllll}
  Set & $am_{b}$ & $aM_{b\overline{b}}$ 
  & $aE_{\eta_b}$ & $aE_{\Upsilon}$
  & $am_s$ & $aM_{\eta_s}$
   & $aE_{B_s}$  & $a\Delta^{hyp}_s$ & $a\Delta^{0^+-0^-}_{B_s}$ & $a\Delta^{1^+-1^-}_{B_s}$ \\ 
\hline
1 & 3.297 & 7.119(9) & 0.21289(6) & 0.26420(8) & 0.0641 & 0.51491(14) & 0.61558(47) & 0.03892(40) & 0.282(12) & 0.289(17)\\ 
2 & 3.263 & 7.040(8) & 0.21546(3) & 0.26669(5) & 0.0636 & 0.51078(8) & 0.61132(26) & 0.03705(47) & 0.285(5) & 0.280(8)\\ 
3 & 2.66 & 5.761(14) & 0.22040(5) & 0.26394(7) & 0.0522 & 0.42351(9) & 0.52385(23) & 0.03177(18) & 0.228(3) & 0.225(5)\\ 
4 & 2.62 & 5.719(7) & 0.22408(3) & 0.26767(5) & 0.0505 & 0.41476(6) & 0.52029(17) & 0.03102(16) & 0.218(6) & 0.222(4)\\ 
5 & 1.91 & 4.264(11) & 0.21519(2) & 0.24802(2) & 0.0364 & 0.30884(11) & 0.41051(17) & 0.02310(14) & 0.164(5) & 0.161(6)\\ 
\end{tabular}
\end{ruledtabular}
\caption{\label{tab:bsresults} 
Results for energies and kinetic masses in lattice units needed for the determination of the mass of the $B_s$ meson.
The second column gives the $b$ quark mass used on each set. The third to fifth columns are the spin average of the $\Upsilon$ and $\eta_b$ kinetic masses along with the ground state energies, the values for sets 3-5 are taken from \cite{Dowdall:2011wh} and use $c_4=1$. 
It was shown in \cite{Dowdall:2011wh} that the spin averaged kinetic mass does not depend strongly on $c_4$ and since $aM_{b\overline{b}}$ is only used for small tuning adjustments this value is sufficient.
Column 6 gives the strange quark mass used in each run.
Column 7 is the mass of the $\eta_s$ meson at the corresponding strange mass, again taken from \cite{Dowdall:2011wh}, apart from retuning on sets 1 and 2.
The ground state energies of the pseudoscalar $B_s$ are given in column 8 and the hyperfine splitting $\Delta_s^{\rm hyp}= E(B_s^*) - E(B_s)$ in column 9.
Columns 10 and 11 give the values of mass differences between scalar and pseudoscalar and between axial vector and vector respectively.
}
\end{table*}

\begin{table}[ht]
\begin{ruledtabular}
\begin{tabular}{cccccc}
%\hline
%\hline
Set & $\Delta_{B_s}$ (GeV) & $\Delta_{M_{b\bar b}}$ (MeV) & $\Delta_{M_{\eta_s}^2}$ (MeV) & $\delta x_l$ & $\delta x_s$ \\
\hline
1 &0.6558(7)(67) 	& -1.5	& 0.0	& 0.17	& 0.01	 \\
2 &0.6533(4)(67) 	& -0.9	& 0.1   & 0.06	& 0.01	 \\
\hline
3 &0.6432(4)(47)	& 2.0	& 1.2	& 0.16	&-0.04	 \\
4 &0.6471(3)(49) 	& 0.0	& 1.4	& 0.06	&-0.04	 \\
\hline
5 &0.6487(4)(44) 	& -1.2	& 0.0	& 0.16	& 0.02	 \\
%\hline
%\hline
\end{tabular}
\end{ruledtabular}
\caption{\label{tab:Bstune} 
Results for $\Delta_{B_s}$ (the mass difference between the $B_s$ meson 
and the spin average of $\Upsilon$ and $\eta_b$ masses) 
on different ensembles.
The two errors are statistics and lattice spacing uncertainty.
Column 3 and 4 give the shifts in MeV that are applied to $\Delta_{B_s}$ to compensate for the 
mistuning of the $b$ and $s$ quarks respectively. Errors are 50\% of the value given.
 Columns 5 and 6 give $\delta x_l$ and $\delta x_s$, the fractional mistuning of the 
sea quark masses in units of the $s$ quark mass, as defined 
in the text.}
\end{table}

%%%%%%%%%%%%%%%%%%%%%%%%%%%%%%%%%%%%%%%%%%%%%%%%%%%%%%%%%%%%%%%%%
\subsubsection{Extracting physical results}
\label{subsubsec:Bsmeson-fits}
%%%%%%%%%%%%%%%%%%%%%%%%%%%%%%%%%%%%%%%%%%%%%%%%%%%%%%%%%%%%%%%%%

Extracting continuum results from a lattice NRQCD calculation is more complicated than in a relativistic formalism due to the way coefficients scale with the cutoff. Usually, one appropriately tunes parameters in the action so that the results are independent of the cutoff up to some power of $a$, and then fits the remaining dependence.
For example in an $\mathcal{O}(a)$ improved action, the following form would be used:
$$
f(a) = f_{\rm phys}\left( 1 + k_1 ( \Lambda a)^2 + k_2( \Lambda a)^4 + ... \right),
$$
where $\Lambda$ sets the scale and logarithmic terms are generally ignored as they are not distinguishable from powers.

Our results here have discretization errors of the above form from 
the light quark and gluon actions. On top of this, our NRQCD action 
will have discretization errors that could have a mild unphysical dependence 
on $am_b$ over the range of $am_b$ values we are using here (1.9-3.3), 
well within the range of validity of NRQCD as an effective theory. 
The $am_b$ dependence comes from missing radiative corrections 
to discretisation correction terms, those with coefficients $c_5$ and 
$c_6$ in Eq.~(\ref{eq:Hamiltonian}). $\mathcal{O}(\alpha_s)$ corrections 
to these coefficients are included here, so the missing terms 
are $\mathcal{O}(\alpha_s^2)$ and higher.  
To allow for this, we include dependence of the discretisation 
errors on $am_b$ in our fits, using the form: 
\begin{eqnarray}
f(a) &=& f_{{\rm phys}}[1 +\sum_  {j=1}^2 d_j(\Lambda a)^{2j}(1 + d_{jb}\delta x_m + d_{jbb}(\delta x_m)^2) ]. \nn\\ \nonumber 
\end{eqnarray} 
Here we model the $am_b$ dependence with a polynomial using 
the parameter $\delta x_m = (am_b-2.7)/1.5$ which varies from approximately 
-0.5 to 0.5 across the range of $am_b$ we use. 
In this way we obtain physical results just as with any other quark 
formalism and the error budget from the fit includes the additional 
error from the effective field theory cutoff dependence.
Note that the effect of relativistic corrections 
to the NRQCD action, which are physical, 
cannot be judged from fitting the data and are included as a separate 
error item.

In practice we find that most quantities in this work have very small lattice spacing dependence. The quantities which do show some dependence are the $B_c$ mass and hyperfine splitting where we believe that the discretisation errors come mainly from the charm quark.

The complete fit function for $\Delta_{B_s}$ also includes terms to allow for sea quark mass dependence.
We take a polynomial in the variables $\delta x_s$ and $\delta x_l$, defined as the difference from the correct quark mass $m_{\rm q,sea,phys}$ normalised by the correct $s$ quark mass
$$
\delta x_q = \frac{\rm m_{q,sea} - m_{\rm q,sea,phys}}{m_{\rm s,sea,phys}}
$$
The values of $\delta x_q$ entering the fits are given in Table \ref{tab:Bstune}.
The values of $\delta x_s$ are significantly smaller than for the 
Asqtad 2+1 ensembles used before~\cite{Gregory:2010gm} and the $\delta x_l$ 
values correspondingly closer to the physical point. 

With this chiral dependence included, the fit function becomes:
\begin{multline}
\label{eq:fitbsextrap}
\Delta_{B_s}(a,\delta x_l, \delta x_s) = \Delta_{B_s,{\rm phys}}[1 \\  
+\sum_{j=1}^2 d_j(\Lambda a)^{2j}(1 + d_{jb}\delta x_m + d_{jbb}(\delta x_m)^2) \\  
+ 2b_l\delta x_l(1+d_l(\Lambda a)^2) \\ 
+ b_s\delta x_s(1+d_s(\Lambda a)^2) \\ 
+ 4b_{ll}(\delta x_l)^2 + 2b_{ls}\delta x_l\delta x_s + b_{ss}(\delta x_s)^2].
\end{multline} 

We take the prior on $\Delta_{B_s,{\rm phys}}$ to be 0.6(2) and we take the physical scale to be $\Lambda=400$ MeV based on the typical meson momenta.  The other terms and priors are:
\begin{itemize}
 \item The quadratic $a$ dependence terms $d_1,d_l,d_s$ should be $\mathcal{O}(\alpha_s)$ or smaller and so have a prior 0.0(3). 
 \item The leading sea quark mass dependence terms $b_l,b_s$ have priors 0.00(7) since sea quark mass dependence is typically $1/3$ of valence mass dependence which would 
give a slope of 0.2 here. 
 \item Quadratic sea quark mass dependence terms $b_{ll},b_{ls},b_{ss}$ are smaller by another factor of 0.2, giving 0.000(13).
 \item The remaining $a^4$ and $am_b$ terms, $d_2, d_{jb}$ and $d_{jbb}$, are given a wide prior of 0(1). 
\end{itemize}
The fit gives $\Delta_{B_s,{\rm phys}}=0.644(6)$ GeV and is robust under changes in the priors and fit function. 
The 6 MeV error can be broken down into contributions from 
$a$ dependence, sea quark mass dependence and the error on the 
data points by looking at the variation of the $\chi^2$ \cite{fitcode}. 
These contributions are listed separately in our final error budget and are dominated by the error on the data points, i.e. statistics and lattice spacing uncertainty. 
Since the quark masses are very well tuned, the corrections for mistuning 
applied in the previous section produce negligible effects.

%%%%%%%%%%%%%%%%%%%%%%%%%%%%%%%%%%%%%%%%%%%%%%%%%%%%%%%%%%%%%%%%%
\subsubsection{Systematic errors}
\label{subsubsec:Bsmeson-syst}
%%%%%%%%%%%%%%%%%%%%%%%%%%%%%%%%%%%%%%%%%%%%%%%%%%%%%%%%%%%%%%%%%
We now describe the  remaining sources of systematic error that cannot be estimated from the fit. The largest of these is the spin independent NRQCD systematic error although there is a significant improvement over previous work due to the inclusion of radiative corrections.

\paragraph*{Spin independent NRQCD systematics:} This error can affect both the bottomonium and $B_s$ pieces of $\Delta_{B_s,{\rm phys}}$. For bottomonium, the NRQCD action is correct through $\mathcal{O}(\alpha_s v^4)$ so the largest errors will be $\alpha_s^2 v^4$ and $v^6$. $v^2$ effects are of order 500 MeV, so we allow an error of $0.3^2 \times 0.1 \times 500=4.5$ MeV from missing $\alpha_s v^4$ corrections (compared to 15 MeV in \cite{Gregory:2010gm}.) Similarly, $v^6$ terms should be 5 MeV. Adding these in quadrature and dividing by two gives 3.4 MeV. For the $B_s$, power counting is in terms of $v= \Lambda/m_b$ which is even smaller in a heavy-light meson and missing spin independent corrections are negligible.

\paragraph*{Spin dependent NRQCD systematics:} Since the bottomonium energies are spin averaged, the only contribution from spin dependent terms is to the $B_s$ mass. With the one loop corrections to $c_4$, the dominant error comes from radiative corrections to the $\sigma.B$ term  and missing $(\Lambda/m_b)^2$ terms. We take the error to be $3\alpha_s^2/4$ times the hyperfine splitting $B_s^* - B_s$ which gives 3 MeV. 

\paragraph*{Electromagnetism:} The effects of missing electromagnetism were 
estimated in \cite{Gregory:2010gm} and give a 0.1 MeV error in the $B_s$. 
\paragraph*{Finite volume effects:} Chiral perturbation theory and studies of the wave functions of heavy mesons show that finite volume errors are negligible for the ensembles used here.
\paragraph*{$M_{\eta_s}$ and $M_{b \bar b}$:} The uncertainty in the $\eta_s$ mass and the error from the electromagnetic and annihilation corrections in $M_{b \bar b}$ also feed into the total error. $M_{\eta_s}$ has an error of 1.2 MeV which using the slope of 0.24 vs $M_{\eta_s}^2$ gives an error of 0.4 MeV to be added to $\Delta_{B_s}$. The error in the adjusted value of $M_{b \bar b}=9.445(2)$ GeV has negligible effect on $\Delta_{B_s}$, but when reconstructing $M_{B_s}$ this leads to a 1 MeV error. The error in $M_{b \bar b}$ comes entirely from electromagnetism/annihilation as the experimental error is negligible.

\begin{figure}
\includegraphics[width=0.9\hsize]{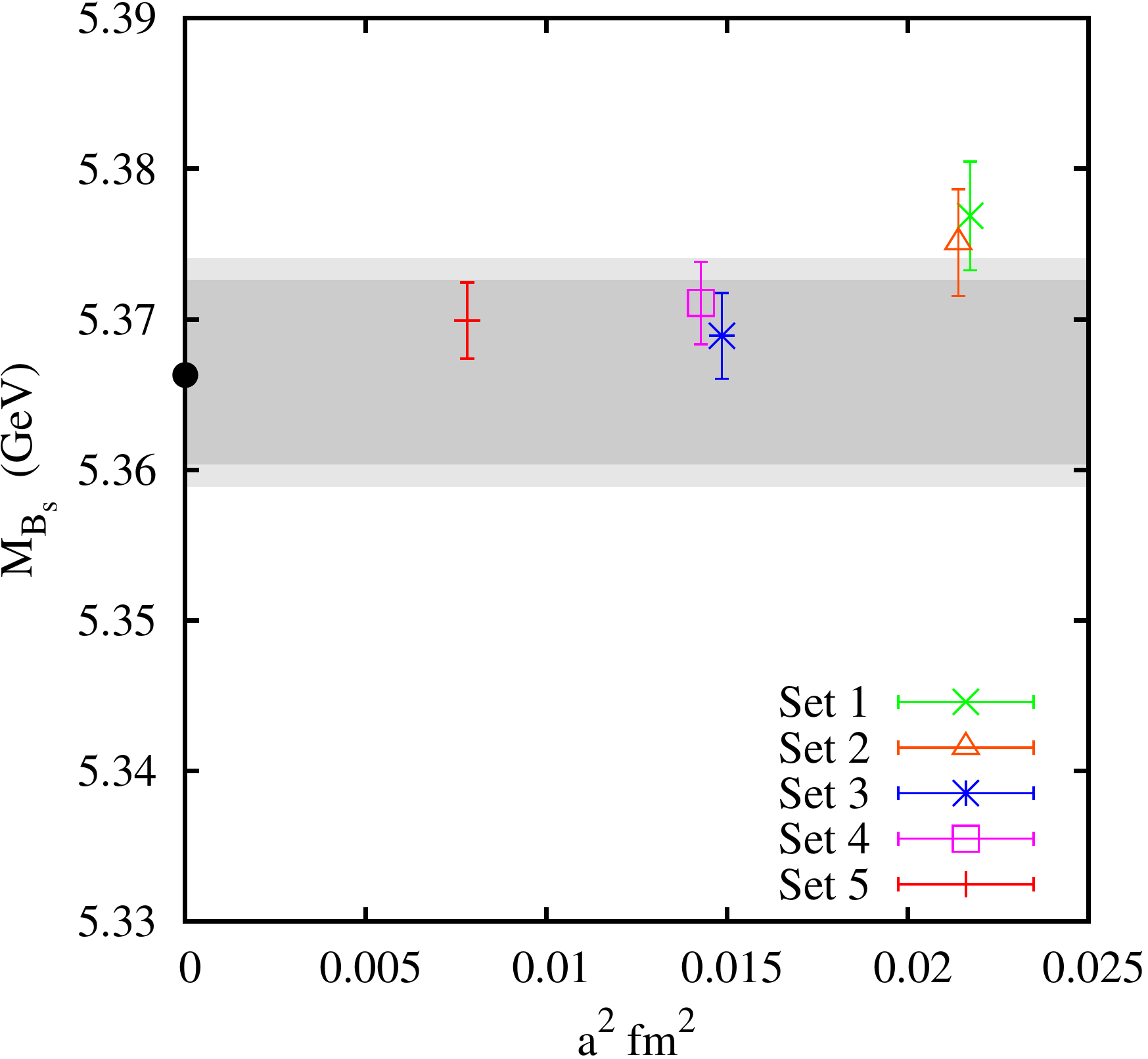}
\caption{Results for the $B_s$ meson mass for each ensemble against the lattice spacing. 
The darker shaded band shows the 6 MeV error from the fit and the light band includes the estimate of systematic errors. Error bars on the data points are uncorrelated and come from statistics, quark mass mistuning and uncertainty in the lattice spacing. The data points 
have been corrected for mistuning of valence quark masses and missing electromagnetism/$\eta_b$ annihilation effects. The experimental value is included in black for comparison.}
\label{fig:Bs}
\end{figure}

\begin{table}[ht]
\begin{ruledtabular}
\begin{tabular}{lccc}
%\hline
%\hline
Error 				& $M_{B_s}$ & $M_{B_c, hh}$ & $M_{B_c, hs}$ \\
\hline
Stats/tuning/uncty in $a$  	& 4.8 	& 2.1 	& 9.5 	\\
Lattice spacing dependence 	& 2.2 	& 0.6 	& 3.5 	\\
$m_{b}$ dependence 		& 2.8 	& 2.9 	& 3.5  	\\
$m_{q,sea}$ dependence 		& 1.4 	& 1.4 	& 5.0  	\\
spin-ind. NRQCD systs. 		& 3.4 	& 5.3 	& 2.3	\\
spin-dep. NRQCD systs. 		& 3.0 	& 3.0 	& 0.0 	\\
uncty in $M_{\eta_s}$ 		& 0.3 	& - 	& 0.7	\\
em, annihiln in $b\overline{b}$ & 1.0   & $1.0^*$ & 0.0	\\
em, annihiln in $c\overline{c}$ & - 	& $1.5^*$ & 0.2 \\
em effects in $B_s$ or $B_c$ 	& 0.1 	& $1.0^*$ & 1.0 \\
em effects in $D_s$  		& - 	& - 	& 1.0 	\\
finite volume 			& 0.0 	& 0.0 	& 0.0 	\\
\hline
Total (MeV) 			& 7.7	& 8.0	& 12.2 \\
%\hline
%\hline
\end{tabular}
\end{ruledtabular}
\caption{\label{tab:errors} 
Full error budget 
for $B_s$ and $B_c$ meson masses in MeV. 
The source of each error is described in the text and the total error is
obtained by adding in quadrature.
Starred errors are correlated and are added linearly before 
being squared.
 }
\end{table}
The systematic errors are summarised in the error budget in Table \ref{tab:errors}. When added in quadrature the total systematic error is 4.7 MeV giving a final value of 
$$M_{B_s} = 5.366(6)(5) {\rm \ GeV,} $$
which should be compared with the current PDG value of 5.3668(2) MeV~\cite{pdg}. 
This is the best result for this quantity from lattice QCD so far. 
There is a noticable improvement over the systematic errors in Ref.~\cite{Gregory:2010gm} but the lattice spacing uncertainty remains similar. The results are plotted in Fig. \ref{fig:Bs}.

%%%%%%%%%%%%%%%%%%%%%%%%%%%%%%%%%%%%%%%%%%%%%%%%%%%%%%%%%%%%%%%%%
%
\subsection{The $D_s$ meson}
\label{subsec:Dsmeson}
%
%%%%%%%%%%%%%%%%%%%%%%%%%%%%%%%%%%%%%%%%%%%%%%%%%%%%%%%%%%%%%%%%%

\begin{table*}
\begin{ruledtabular}
\begin{tabular}{lllllllllll}
Set & $am_{c}$ & $am_{s}$ & $am_l$ &
$aE_{\eta_c}$ & $aE_{D_s}$ & $aE_{D}$ &
$aE_{D_s} - aE_{D}$  & $ aE_{D_s} - aE_{\eta_c}/2 $  
& $\Delta_{M_{\eta_s}^2 }$ &  $\Delta_{M_{\eta_c}}$
\\
\hline
1 & 0.826 & 0.0641 & 0.013 & 2.22508(7) & 1.48729(30) & 1.43326(58) & 0.05403(57) & 0.37475(29)&0.0&0.3\\ 
2 & 0.818 & 0.0636 & 0.0064 & 2.21032(4) & 1.47559(20) & 1.41258(68) & 0.06300(69) & 0.37043(19)&0.0&0.2\\ 
3 & 0.645 & 0.0522 & 0.01044 & 1.83967(5) & 1.21934(14) & 1.17112(53) & 0.04822(47) & 0.29950(13)&0.7&0.4\\ 
4 & 0.627 & 0.0505 & 0.00507 & 1.80351(3) & 1.19554(8) & 1.14112(61) & 0.05442(58) & 0.29379(8)&0.8&0.4\\ 
5 & 0.434 & 0.0364 & 0.0074 & 1.33307(4) & 0.88212(9) & 0.84682(26) & 0.03530(21) & 0.21559(10)&0.0&0.4\\ 
\end{tabular}
\end{ruledtabular}
\caption{\label{tab:Dresults} 
Results from charmed meson fits in lattice units. Columns 2-4 give the HISQ quark masses used in the run, columns 5-7 give the $\eta_c$, $D_s$ and $D$ energies with statistical errors only. Columns 8 and 9 give splittings that we use in our fits. They have reduced errors over the naive subtraction of earlier columns because correlations are taken into account.
Columns 10 and 11 give the shifts in MeV that are applied to $\Delta_{D_s}$ due to mistuning of the $s$ and $c$ quarks.
}
\end{table*}

Our method for calculating the mass of the $D_s$ meson closely follows that of \cite{newfds}. The previous study on MILC 2+1 AsqTad ensembles included 5 values of the 
lattice spacing down to 0.045 fm and found $M_{D_s}=1.9691(32)$ GeV. 
Here we have only  3 lattice spacings at the coarser end of the range so 
our result will suffer from a larger error from the continuum extrapolation. 
Some other systematic errors are smaller, however and 
our results provide an interesting comparison with those in the $B$ spectrum.

To determine $M_{D_s}$, we calculate the splitting $M_{D_s}-M_{\eta_c}/2$ which has several advantages over determining the mass directly. Since the splitting is much smaller than the mass, the same relative scale uncertainty translates into a much smaller absolute error on the splitting. It was shown in \cite{newfds} that the $c$ quark mass dependence of the splitting is small which leads to reduced tuning errors, particularly on the coarsest ensembles where discretisation errors are large. 
Finally, the splitting allows for a direct comparison with $M_{B_s}-M_{\eta_b}/2$ which must be used in the NRQCD case due to the unphysical energy shift. The $\eta_c$ is used rather than the spin averaged $c \bar c$ state simply because a staggered vector meson would require additional propagators to be generated.  
$M_{B_s}-M_{\eta_b}/2$ has a slightly increased systematic error over our 
preferred $\Delta_{B_s}$ (eq.~(\ref{EXPI})).

Like the NRQCD $B_s$ correlators, the $D_s$ fit function includes oscillating terms coming from the states related by parity and, being relativistic, also includes cosh time-dependence:
\begin{eqnarray}
\label{eq:Dsfitform}
C_{\rm meson}(t)&=& 
\sum^{N_{\exp}}_{k=1} a_k \left( e^{-E_{k}t} + e^{-E_{k}(T-t)} \right)
\nn \\
&-& (-1)^{t}\sum^{N_{\exp}-1}_{k^\prime=1}   b_{k^\prime}
\left( e^{-E^\prime_{k^\prime}t} + e^{-E^\prime_{k^\prime}(T-t)}
\right)
\end{eqnarray}
As for the $B_s$ fits, the priors on the energy splittings $E_{n+1}-E_n$ are taken to be approximately 600(300) MeV and the prior on the ground state is 1.9 GeV with a 300 MeV width. Similarly the prior splitting between the ground state and first oscillating state is 600(300) MeV.
Fits with $N_{\exp}=5$ are typically used as the results are stable by this point. 
$D_s, D$ and $\eta_c$ correlators are fit simultaneously on each ensemble to include the correlations in the splittings.

Before performing a continuum extrapolation we must correct for mistuning of the valence quark masses. The $s$ and $c$ quark mass dependence of $M_{D_s}-M_{\eta_c}/2$ was studied in detail in Ref.~\cite{newfds} by fitting the splitting as a function of $M_{\eta_s}^2$ and $M_{\eta_c}$.
The dependence is linear over the range of values used with a slope of 0.20(1) against $M_{\eta_s}^2$ and 0.05 against $M_{\eta_c}$.
Although this data used AsqTad sea quarks the corrections are small and since all shifts are applied with a 50\% error, any difference between the slope for HISQ sea quarks will be negligible.
The shifts applied to $M_{D_s}-M_{\eta_c}/2$ are listed in Table \ref{tab:Dresults}.
Another advantage of using $M_{D_s}-M_{\eta_c}/2$ is that the error from the lattice spacing is a third of the naive value. Changing the lattice spacing requires $m_s$ and $m_c$ to be retuned, the effect of which partially cancels in the splitting.

The results at different lattice spacings and light quark masses are fit to the same function as in Ref.~\cite{newfds}
\begin{multline}
\label{eq:fitDsextrap}
\Delta_{D_s}(a,\delta x_l, \delta x_s) = 
\Delta_{D_s,{\rm phys}}[1  
+\sum_{j=1}^4 d_j(m_c a)^{2j}  \\
+ 2b_l\delta x_l(1+d_l(m_c a)^2)  
+ b_s\delta x_s(1+d_s(m_c a)^2) \\ 
+ 4b_{ll}(\delta x_l)^2 + 2b_{ls}\delta x_l\delta x_s + b_{ss}(\delta x_s)^2].
\end{multline} 
The same prior values as for the $B_s$ are used for the sea quark mass dependence and 
the splitting itself is taken to have prior 0.5(2) GeV. The discretisation terms have priors 0.0(2) except for $d_1$ which is 0.00(6) since tree-level $a^2$ errors have been removed in the HISQ action. Discretisation errors are set by the scale $\Lambda=m_c$ since the dominant error will come from the charm quarks.

\begin{figure}
\includegraphics[width=0.9\hsize]{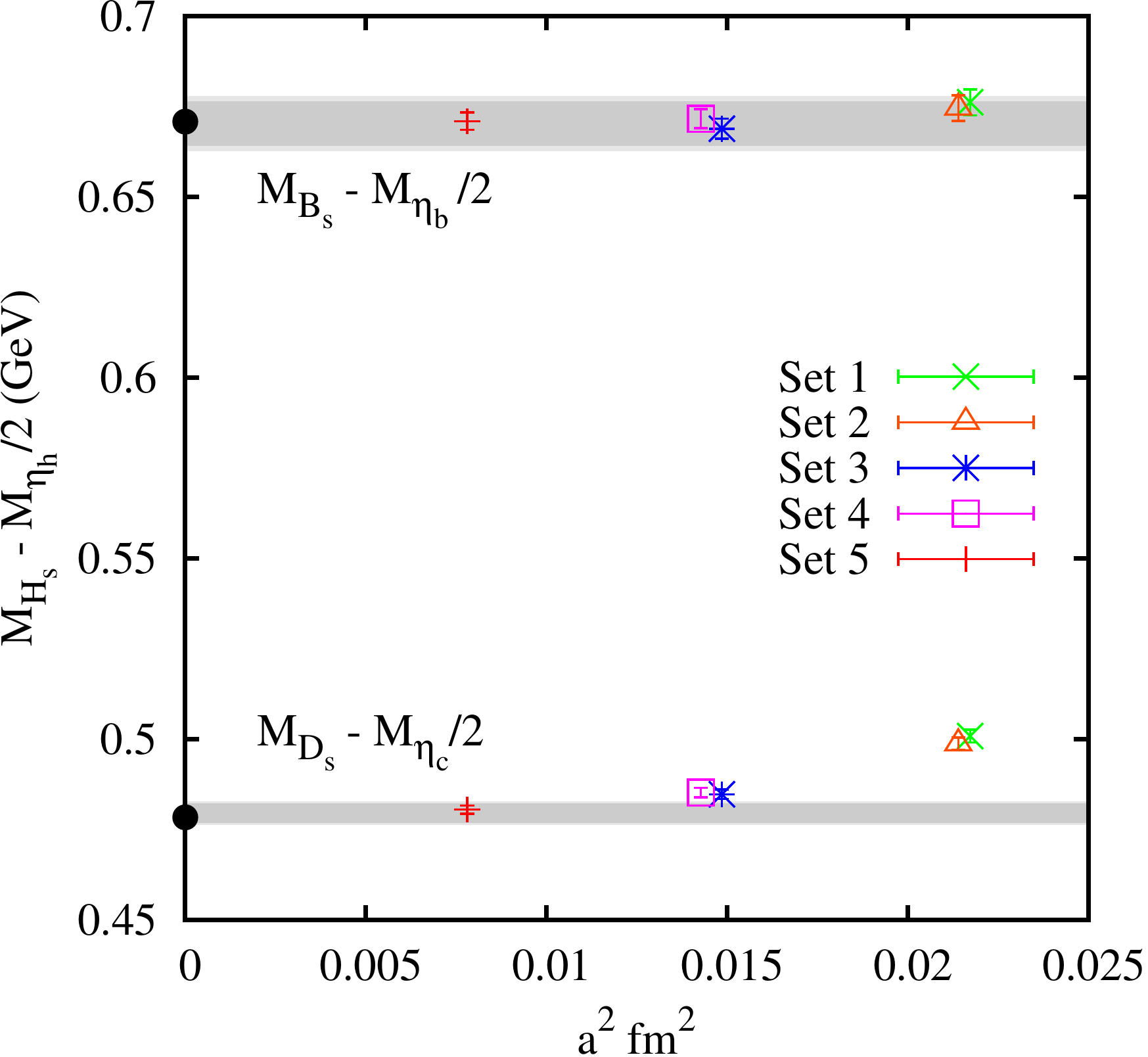}
\caption{
Plot of the splittings $M_{D_s}-M_{\eta_c}/2$ and $M_{B_s}-M_{\eta_b}/2$ against the 
square of the lattice spacing.
The dark grey band is the statistical error and the lighter band gives the full combined statistical and systematic error. The errorbars include statistics, scale and tuning only, correlated errors are not shown. The lattice results are adjusted for missing electromagnetic 
and annihilation effects.
}
\label{fig:Ds-etac}
\end{figure}

The result of the fit, 0.4808(28) GeV, is plotted in Fig.~\ref{fig:Ds-etac} 
along with the retuned data on each ensemble. Also included for comparison is the corresponding splitting in the $B$ meson spectrum $M_{B_s}-M_{\eta_b}/2$ from our results.
There is a significant difference in the two splittings, largely driven by 
the stronger binding of heavyonium as the heavy quark mass is increased. The experimental 
difference between $c$ and $b$ is well reproduced by our results here. 
The complete dependence on heavy quark mass is mapped out in~\cite{McNeile:2011ng, McNeile:2012qf}. 
The lighter shaded band in Fig.~\ref{fig:Ds-etac} includes the systematic errors which are discussed in the next section.

%%%%%%%%%%%%%%%%%%%%%%%%%%%%%%%%%%%%%%%%%%%%%%%%%%%%%%%%%%%%%%%%%
\subsubsection{Systematic errors}
\label{subsubsec:Dsmeson-syst}
%%%%%%%%%%%%%%%%%%%%%%%%%%%%%%%%%%%%%%%%%%%%%%%%%%%%%%%%%%%%%%%%%

The error arising from statistical/scale, lattice spacing dependence and sea quark mass effects is estimated from the fit as above, the remaining systematic errors that cannot be found in this way are the following:
\paragraph*{Electromagnetism:} Electromagnetic effects in the $D_s$  were estimated in Ref.~\cite{newfds} where the shift was 1.3(7) MeV, assuming a 50\% error.
\paragraph*{$M_{\eta_s}$:} The uncertainty in the mass of the $\eta_s$ meson, used for tuning to the correct $s$ quark mass, feeds into the error. Using the slope of 0.2, the mass $M_{\eta_s}=0.6893(12)$ GeV results in an error of $0.2\times2\times 1.2\times 0.69\sim0.3$ MeV.
\paragraph*{$M_{\eta_c}$:} When the $D_s$ mass is reconstructed from the splitting, we must include the error from $M_{\eta_c}=2.985(3)$ that comes from our estimate of electromagnetic and annihilation effects in the $\eta_c$ as well as experimental errors. 
This gives a 1.5 MeV error in $M_{D_s}$
\paragraph*{Lattice spacing systematics:} Systematic errors in the determination of the lattice spacing are included in the scale error.

The error budget for $M_{D_s}$ is given in Table~\ref{tab:Dserrors} and our final result is
$$
M_{D_s} = 1.9697(28)(17) {\ \rm GeV}
$$
where the two errors are fitting/scale/tuning and systematics and lead to a combined error of 3.3 MeV.
In fact our final error is not significantly worse than in~\cite{newfds} because 
an increased lattice spacing extrapolation error is offset by the accurate 
physical value for $M_{\eta_s}$. The current experimental result for 
$M_{D_s}$ is 1.9685(3) GeV~\cite{pdg}. 

\begin{table}[ht]
\begin{ruledtabular}
\begin{tabular}{lc}
%\hline
%\hline
Error 				& $M_{D_s}$   \\
\hline
Stats/tuning/uncty in $a$  	& 2.2 	  \\
Lattice spacing dependence 	& 1.6 	  \\
$m_{q,sea}$ dependence 		& 0.7 	  \\
uncty in $M_{\eta_s}$ 		& 0.7	  \\

em effects in $D_s$ 		& 0.7   \\
em, annihiln effects in $\eta_c$& 1.5 	 \\
finite volume 			& 0.0 	 \\
\hline
Total (MeV) 			& 3.3	 \\
%\hline
%\hline
\end{tabular}
\end{ruledtabular}
\caption{\label{tab:Dserrors} 
Full error budget for $M_{D_s}$ in MeV. The different errors are described in more detail in the text.
 }
\end{table}

%%%%%%%%%%%%%%%%%%%%%%%%%%%%%%%%%%%%%%%%%%%%%%%%%%%%%%%%%%%%%%%%%
%
\subsection{The $B_c$ meson}
\label{subsec:Bcmeson}
%
%%%%%%%%%%%%%%%%%%%%%%%%%%%%%%%%%%%%%%%%%%%%%%%%%%%%%%%%%%%%%%%%%

\begin{table*}
\begin{ruledtabular}
\begin{tabular}{llllllllllll}
Set & $am_{b}$ & $am_{c}$ & $am_s$ &
$aE_{\eta_c}$ & $aE_{D_s}$ &
$aE_{B_c}$  & $a\Delta^{\rm hyp}_{B_c}$ &   $a\Delta^{0^+-0^-}_{B_c}$ &  $a\Delta^{1^+-1^-}_{B_c}$
&$aE_{B_c'} - aE_{B_c}$ & $aE_{B_c^{*'}} - aE_{B_c^*}$
\\
\hline
1 & 3.297 & 0.826 & 0.0641 & 2.22508(7) & 1.48729(30) & 1.30409(14) & 0.03659(17) & 0.256(87) & 0.212(73) &  - &  - \\ 
2 & 3.263 & 0.818 & 0.0636 & 2.21032(4) & 1.47559(20) & 1.29702(10) & 0.03658(13) & 0.335(27) & 0.327(39) &  - &  - \\ 
3 & 2.66 & 0.645 & 0.0522 & 1.83967(5) & 1.21934(14) & 1.08866(5) & 0.03140(3) & 0.241(17) & 0.250(9) & 0.618(27) & 0.605(19)\\ 
4 & 2.62 & 0.627 & 0.0505 & 1.80351(3) & 1.19554(8) & 1.07252(4) & 0.03137(2) & 0.252(5) & 0.242(6) & 0.637(15) & 0.625(13)\\ 
5 & 1.91 & 0.434 & 0.0364 & 1.33307(4) & 0.88212(9) & 0.81480(3) & 0.02470(2) & 0.190(2) & 0.184(2) & 0.616(8) & 0.591(7)\\ 
\end{tabular}
\end{ruledtabular}
\caption{\label{tab:bcresults}
Parameters and results of the $B_c$ meson mass calculations.
The first three columns give the bottom, charm and strange quark masses used in the runs.
$aM_{\eta_c}, aM_{D_s}$ are the masses of the pseudoscalar mesons generated with the same HISQ propagators used for the $B_c$ and $B_s$. 
$aE_{B_c}$ and $a\Delta^{\rm hyp}_{B_c}$ are the ground state energy and hyperfine splitting on each ensemble.
Columns 9 and 10 give the splittings with the parity partner states discussed in Sec.~\ref{sec:axialmesons}.
The final two columns give radial excitation energies for the $B_c$ and $B_c^*$. 
}
\end{table*}

In Ref.~\cite{Gregory:2010gm}, two different methods of reconstructing the $B_c$ mass were used: the ``heavy-heavy'' (or hh) subtraction method and the ``heavy-strange'' (or hs) subtraction. In the hh method, half the mass of the $\eta_c$ is subtracted from the lattice value of $E_{B_c}$ in addition to the spin averaged bottomonium ground state energy.
\begin{eqnarray}
M_{B_c} &=& \left(aE_{B_c} -\frac{1}{2}(aE_{b\overline{b}}
+aM_{\eta_c})\right)_{\rm latt}a^{-1} \nonumber \\ 
&+& \frac{1}{2}\left(M_{b\overline{b},{\rm phys}} + M_{\eta_c,{\rm phys}}\right).
\label{hh-method}
\end{eqnarray}
This has two advantages, firstly it makes the splitting a very small value 
which results in a reduced error from the uncertainty in the 
lattice spacing, and secondly it reduces mistuning errors since to a good approximation $E_{B_c}$ and $M_{\eta_c}$ depend linearly on the charm quark mass.
The second method, hs, uses the $B_s$ and $D_s$ energies to remove the unphysical energy shift from NRQCD:
\begin{eqnarray}
\label{hs-method}
M_{B_c} &=& \left(aE_{B_c} - (aE_{B_s}+aM_{D_s})\right)_{\rm latt}a^{-1}  \nonumber \\
&+& \left(M_{B_s, {\rm phys}} + M_{D_s, {\rm phys}}\right).
\end{eqnarray}
The $D_s$ and $\eta_c$ masses are calculated using HISQ for both the $c$ and $s$ valence quarks with the parameters given in Table~\ref{tab:hisqparams}.
The hh and hs methods have different systematic errors and give two 
independent results to check consistency. Previously~\cite{Gregory:2010gm} the hh and hs methods resulted in total 
errors in the $B_c$ mass of 10 MeV and 19 MeV respectively, using NRQCD $b$ quarks. 

Table \ref{tab:bcresults} gives the energies of the $B_c$, $D_s$ and $\eta_c$ required for the two methods. The $B_s$ energies are those given in Table~\ref{tab:bsresults}.

%%%%%%%%%%%%%%%%%%%%%%%%%%%%%%%%%%%%%%%%%%%%%%%%%%%%%%%%%%%%%%%%%
\subsubsection{Heavy-heavy method}
\label{subsubsec:Bc-hh}
%%%%%%%%%%%%%%%%%%%%%%%%%%%%%%%%%%%%%%%%%%%%%%%%%%%%%%%%%%%%%%%%%

We begin with the hh method, values for $\Delta_{B_c,hh}$ are listed in Table \ref{tab:tunebc}. As for the $B_s$ we need to correct $\Delta_{B_c,hh}$ for small mistunings in the quark masses.
In \cite{Gregory:2010gm} the slope with respect to  $M_{b\overline{b}}$ was 
0.014 (agreeing with that from using HISQ $b$ quarks in~\cite{McNeile:2012qf}) 
which gives us the shifts $\Delta_{M_{b\bar b},hh}$ given in Table \ref{tab:tunebc}.
The shifts are around 1 MeV which is comparable to, or slightly larger than, 
the lattice spacing uncertainty. 
Since the slope is a physical dependence rather than a lattice artefact, for the charm quark we use the slope against $M_{\eta_c}$ of -0.035 found in \cite{McNeile:2012qf}. 
This was based on more data and on finer lattice spacings than the 
smaller value in~\cite{Gregory:2010gm}.
From this we obtain the shifts, $\Delta_{M_{\eta_c},hh}$, in Table~\ref{tab:tunebc}.
These shifts are negligible compared to the lattice spacing errors. 
Again, the errors on the shifts are taken to be 50\% of the shift.
As for the $B_s$, once retuning is taken into account the actual scale error on the splitting is less than the naive value, in this case ranging from 0.5-0.7 of the naive value. We take 0.7 times the $a$ error on all ensembles.

\begin{table}[ht]
\begin{ruledtabular}
\begin{tabular}{cccc}
%\hline
%\hline
Set & $\Delta_{B_c,hh}$ (GeV) & $\Delta_{M_{b\bar b},hh}$ (MeV)& $\Delta_{M_{\eta_c},hh}$ (MeV)\\
\hline
1 & 0.0882(2)(9) & -1.2 & -0.2 \\
2 & 0.0876(2)(9) & -0.7 & -0.1 \\
\hline
3 & 0.0685(1)(5)   & 1.7  & -0.2 \\
4 & 0.06999(7)(52) & 0.02 & -0.2\\
\hline
5 & 0.06331(9)(43) & -1.0 & -0.3\\
%\hline
%\hline
\end{tabular}
\end{ruledtabular}
\caption{\label{tab:tunebc} 
Table shows the energy splittings $\Delta_{B_c,hh}$ where the two errors are statistics and lattice spacing uncertainty. The second and third columns are the shifts in MeV applied to $\Delta_{B_c,hh}$ to adjust for $b$ and $c$ quark mass mistuning respectively. 
}
\end{table}

The data are fit to a similar form to that of $\Delta_{B_s}$ but with a few changes. 
Since $\Delta_{B_c,hh}$ has such a small value, the scale, cutoff and 
sea quark mass dependence are included additively rather than multiplicatively 
to allow them a larger range. 
We give them instead an overall coefficient of 0.4 GeV. 
We also expect the discretisation errors to be dominated by the charm quark so $m_c \simeq 1$ GeV is used instead of $\Lambda$ to set their scale.
Our fit form is then:
\begin{eqnarray}
\label{eq:fitbcextrap}
\Delta_{B_c,hh}(a,\delta x_l, \delta x_s) &=& \Delta_{B_c,hh,{\rm phys}} + \\ \nonumber 
0.4\big[ \sum_{j=1}^4 d_j(m_c a)^{2j}(&1& + d_{jb}\delta x_m + d_{jbb}(\delta x_m)^2) \\ \nonumber 
+ 2b_l\delta x_l(&1&+d_l(m_c a)^2+d_{ll}(m_c a)^4) \\ \nonumber
+ 2b_s\delta x_s(&1&+d_s(m_c a)^2+d_{ss}(m_c a)^4) \\ \nonumber
+  4b_{ll}(\delta x_l)^2 &+& 2b_{ls}\delta x_l\delta x_s + b_{ss}(\delta x_s)^2\big] .
\end{eqnarray} 
We take the prior on $\Delta_{B_c,hh,{\rm phys}}$ to be 0.05(5). The priors for the fit terms are the same as for the $B_s$ case with the additional $d_j,d_{jb},d_{jbb}$ terms having priors of 0(1).
The fit gives $\Delta_{B_c,hh,{\rm phys}}=0.06131(39)$ (fit error only), the systematic errors that must be included when reconstructing $M_{B_c}$ are the following:

\begin{figure}
\includegraphics[width=0.9\hsize]{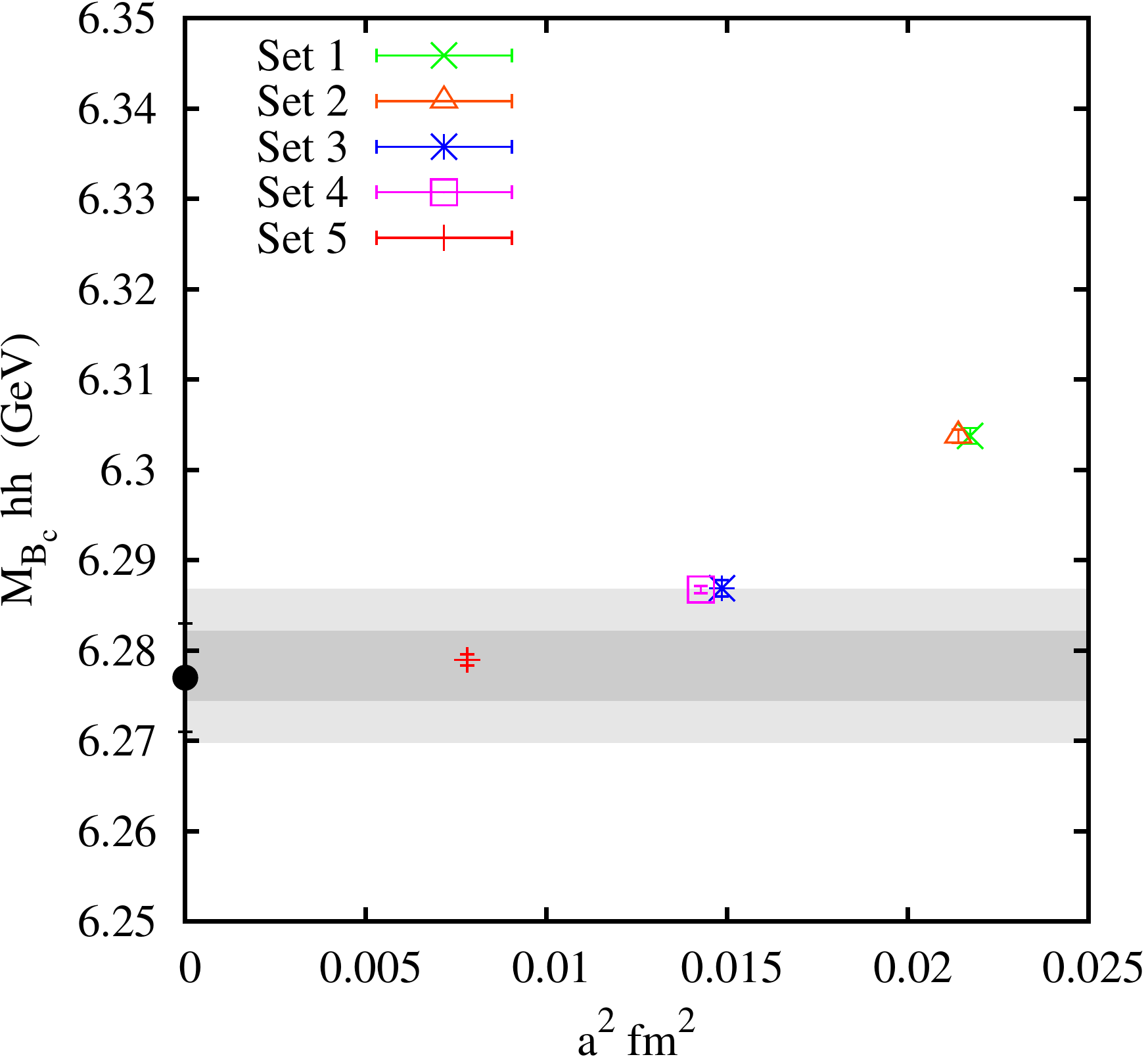}
\caption{Results for the $B_c$ meson mass for each ensemble plotted against the lattice spacing.
The errors on the data points include statistics, fitting and lattice spacing uncertainty and are adjusted for electromagnetic corrections and mistuning of quark masses.
The dark shaded band gives the error coming from the fit and the lighter band includes all systematic errors discussed in the text.
The black circle is the current experimental value. 
}
\label{fig:Bc-hh}
\end{figure}

\paragraph*{Spin independent NRQCD systematics:} The effect of missing terms in the action on $M_{b \bar b}$ is the same as discussed previously, but since the $b$ quark velocity in the $B_c$ is half that in bottomonium we expect partial cancellation of the $\alpha^2 v^4$ errors in $\Delta_{B_c,hh}$. We take 1.7 MeV, which is half the value for the $B_s$ case. The $v^6$ terms are not expected to cancel and results in the same 5 MeV giving a total of 2.6 MeV when added in quadrature and halved.

\paragraph*{Spin dependent NRQCD systematics:}  As for the $B_s$, we take the error to be $3\alpha_s^2/4$ times the hyperfine splitting in the $B_c$ system giving 3 MeV.

\paragraph*{Electromagnetism:} Electromagnetic effects are not negligible in the $B_c$ and the required shift was estimated in \cite{Gregory:2010gm} to be +2(1) MeV.

\paragraph*{Finite volume effects:} Chiral perturbation theory and studies of the wave functions of heavy mesons show that finite volume errors are negligible for the ensembles used here.

\paragraph*{$M_{\eta_c}$ and $M_{b \bar b}$:} Since the slopes of $\Delta_{B_c,hh}$ against these meson masses are very small, the uncertainty in $M_{\eta_c}$ and $M_{b \bar b}$ does not require an additional error to be included. However the errors will appear when $M_{B_c}$ is reconstructed. These errors come from corrections due to electromagnetism and annihilation effects and are correlated since the same method was used to estimate these shifts. Taking half the error on these shifts gives 1 MeV for $M_{b \bar b}$ and 1.5 MeV for  $M_{\eta_c}$. These are added linearly along with the 1 MeV for electromagnetic effects in the $B_c$ described above before being added in quadrature to the other errors. The correlated errors are marked with a $^*$ in Table \ref{tab:errors}.

Taking all of these systematic errors into account, our value for the $B_c$ mass using the hh method is
$$M_{B_c} = 6.278(4)(8) {\ \rm GeV.}$$
The fit result is plotted in Fig.~\ref{fig:Bc-hh} along with the retuned data points for each ensemble.
Table \ref{tab:errors} gives the contribution to the final error of statistics, tuning, scale uncertainty and quark mass dependence. 
Adding the statistical and systematic errors in quadrature gives a total error of 9 MeV which is shown as the lighter shaded band in Fig. \ref{fig:Bc-hh}.
The current experimental value is 6.277(6) GeV~\cite{pdg}. 

%%%%%%%%%%%%%%%%%%%%%%%%%%%%%%%%%%%%%%%%%%%%%%%%%%%%%%%%%%%%%%%%%
\subsubsection{Heavy-strange method}
\label{subsubsec:Bc-hs}
%%%%%%%%%%%%%%%%%%%%%%%%%%%%%%%%%%%%%%%%%%%%%%%%%%%%%%%%%%%%%%%%%

The hs method requires tuning adjustments for the $b,c$ and $s$ quark masses. 
Ref. \cite{Gregory:2010gm} found strong dependence on the $s$ quark but very small dependence on the $b$ and $c$ masses. The slope against $M_{\eta_s}^2$ is 0.41, the slope against $M_{b \bar b}$ is 0.005 and against $M_{\eta_c}$ is 0.07. 
These slopes agree with the results in~\cite{McNeile:2012qf}. 
The resulting shifts are given in Table \ref{tab:tunebc-hs} along with the energy splittings $\Delta_{B_c,hs}$.
The biggest shifts are those for mistuning of the $s$ quark on the coarse 
lattices, but even there the shifts are smaller than the lattice spacing 
uncertainty.

\begin{table}[ht]
\begin{ruledtabular}
\begin{tabular}{ccccc}
%\hline
%\hline
Set & $\Delta_{B_c,hs}$ (GeV) & $\Delta_{M_{b\bar b},hs}$ & $\Delta_{M_{\eta_c},hs}$  & $\Delta_{M_{\eta_s},hs}$\\
  &   &   & (MeV)   &   \\
\hline
1 & -1.069(1)(10) & -0.4 &  0.4 & 0.0 \\
2 & -1.065(1)(10) & -0.25 &  0.3 & 0.2 \\
\hline
3 & -1.059(1)(7)  &  0.6 &  0.5 & 2.1 \\
4 & -1.062(1)(8)  &  0.0 &  0.5 & 2.5 \\
\hline
5 & -1.067(1)(7)  & -0.4&  0.7 & -0.1\\
%\hline
%\hline
\end{tabular}
\end{ruledtabular}
\caption{\label{tab:tunebc-hs} 
Results for the hs splitting $\Delta_{B_c,hs}$ in GeV where the two errors are statistical and scale uncertainty. Columns 3-5 are the shifts in MeV applied for mistuning of the $b$, $c$ and $s$ quark respectively in MeV.
}
\end{table}

The fit function is the same as the hh case but with the dependences included multiplicatively
\begin{eqnarray}
\label{eq:fitbchsextrap}
\Delta_{B_c,hs}(a,\delta x_l, \delta x_s) &=& \Delta_{B_c,hs,{\rm phys}}\big[ 1 + \\ \nonumber 
 \sum_{j=1}^4 d_j(m_c a)^{2j}(&1& + d_{jb}\delta x_m + d_{jbb}(\delta x_m)^2) \\ \nonumber 
+ 2b_l\delta x_l(&1&+d_l(m_c a)^2+d_{ll}(m_c a)^4) \\ \nonumber
+ b_s\delta x_s(&1&+d_s(m_c a)^2+d_{ss}(m_c a)^4) \\ \nonumber
+  4b_{ll}(\delta x_l)^2 &+& 2b_{ls}\delta x_l\delta x_s + b_{ss}(\delta x_s)^2\big] .
\end{eqnarray} 
The prior on $\Delta_{B_c,hs,{\rm phys}}$ is -1.0(2), all other priors are the same as for the hh method.
The fit result is $\Delta_{B_c,hs,{\rm phys}}=-1.071(12)$ where the error is from the fit only. The systematic errors are listed below:

\begin{figure}
\includegraphics[width=0.9\hsize]{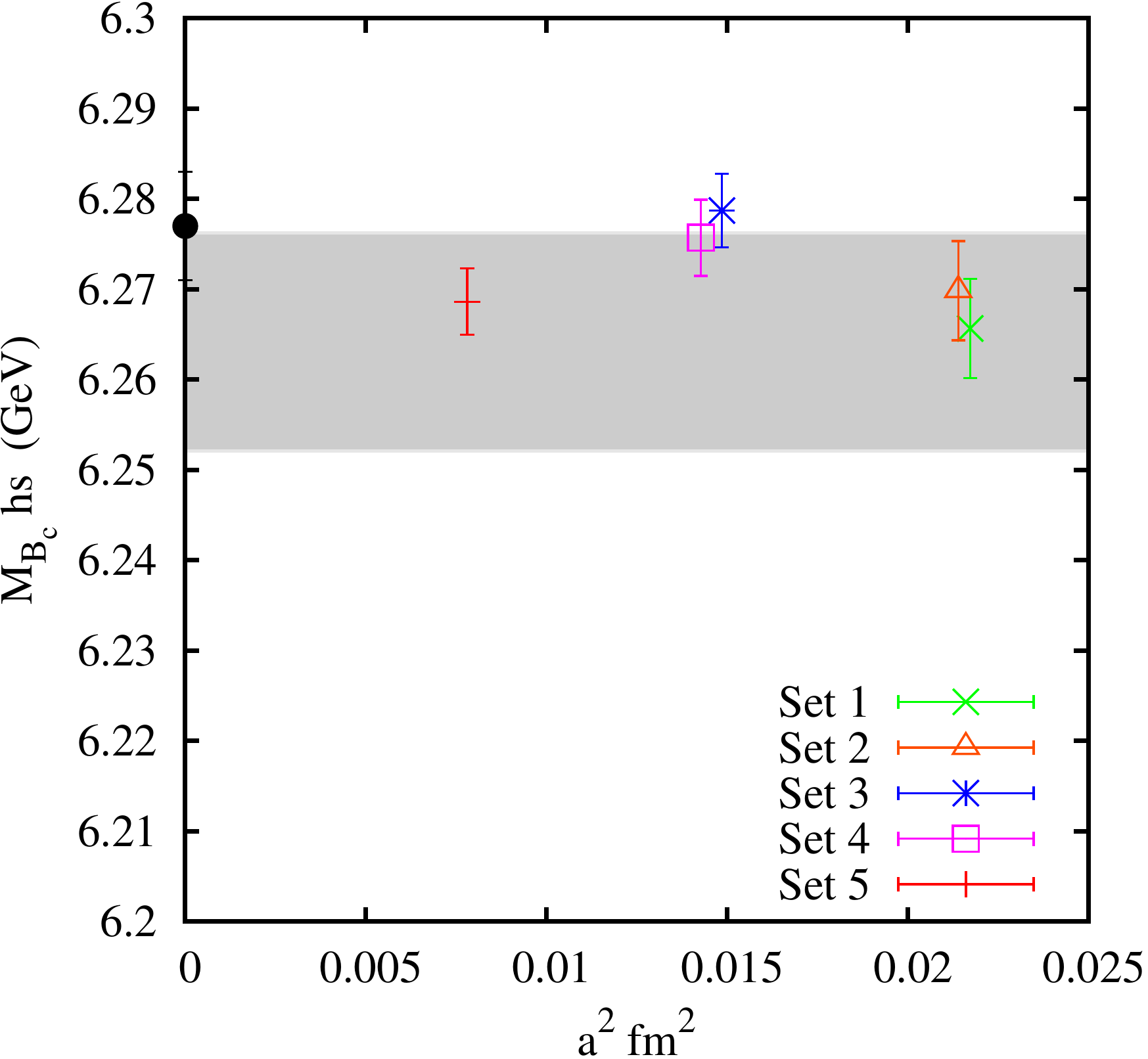}
\caption{Results for the $B_c$ meson mass for each ensemble plotted against the lattice spacing using the hs method. The data points are adjusted for missing electromagnetism and 
mistuning of quark masses.
The dark grey band is the statistical error on the fit result and the systematic error is shown in light grey and is barely visible on this scale.
}
\label{fig:Bc-hs}
\end{figure}

\paragraph*{Spin independent NRQCD systematics:} There will be no cancellation as in the hh case so spin independent systematic errors in the $B_c$ will be of order $\alpha_s^2 v^4$. Based on the $b$ quark velocities in each meson, this should be half as big for the $B_c$ as for $M_{b \bar b}$ estimated earlier, giving 2.3 MeV.
These missing terms also enter the $B_s$ mass but are negligible, along with $v^6$ terms in both mesons.

\paragraph*{Spin dependent NRQCD systematics:}  
The $\sigma\cdot B$ term in the action will affect the $B_s$ and $B_c$ in a similar way 
so errors from unknown $\alpha_s^2$ terms in $c_4$ should be negligible.

\paragraph*{Electromagnetism:} As in the hh method, there is a shift of +2(1) MeV for the $B_c$ but we must also include a shift of -1.3(7) due to the subtraction of the $D_s$ mass. 

\paragraph*{$M_{\eta_s}$, $M_{\eta_c}$ and $M_{b \bar b}$:}
The errors in the retuning coming from the $M_{\eta_c}$ and $M_{b \bar b}$ are negligible due to the small slopes, but the 1.2 MeV error in the $M_{\eta_s}$ results in a 0.7 MeV error in $\Delta_{B_c,hs}$.
We also need to include the error in the reference $B_s$ and $D_s$ masses which is 
dominated by our estimates of electromagnetic corrections. The $D_s$ has an error of 0.7 MeV and it is negligible for the $B_s$.
Including the errors in a correlated way is not necessary here as only the electromagnetic shift in the $B_c$ is not negligible.

Our final answer for the $B_c$ mass with the hs method is 
$$M_{B_c} = 6.264(12)(3) {\rm \ GeV} $$
where the error is dominated by statistics. This is in good 
agreement, but not quite as accurate, as our result from 
the hh method. 
This mass is shown in Fig. \ref{fig:Bc-hs} along with the retuned data points on each ensemble, both corrected for missing electromagnetic effects described above.

%%%%%%%%%%%%%%%%%%%%%%%%%%%%%%%%%%%%%%%%%%%%%%%%%%%%%%%%%%%%%%%%%
\subsubsection{Radially excited states}
\label{subsubsec:Bc-excited}
%%%%%%%%%%%%%%%%%%%%%%%%%%%%%%%%%%%%%%%%%%%%%%%%%%%%%%%%%%%%%%%%%

\begin{figure}
\includegraphics[width=0.75\hsize]{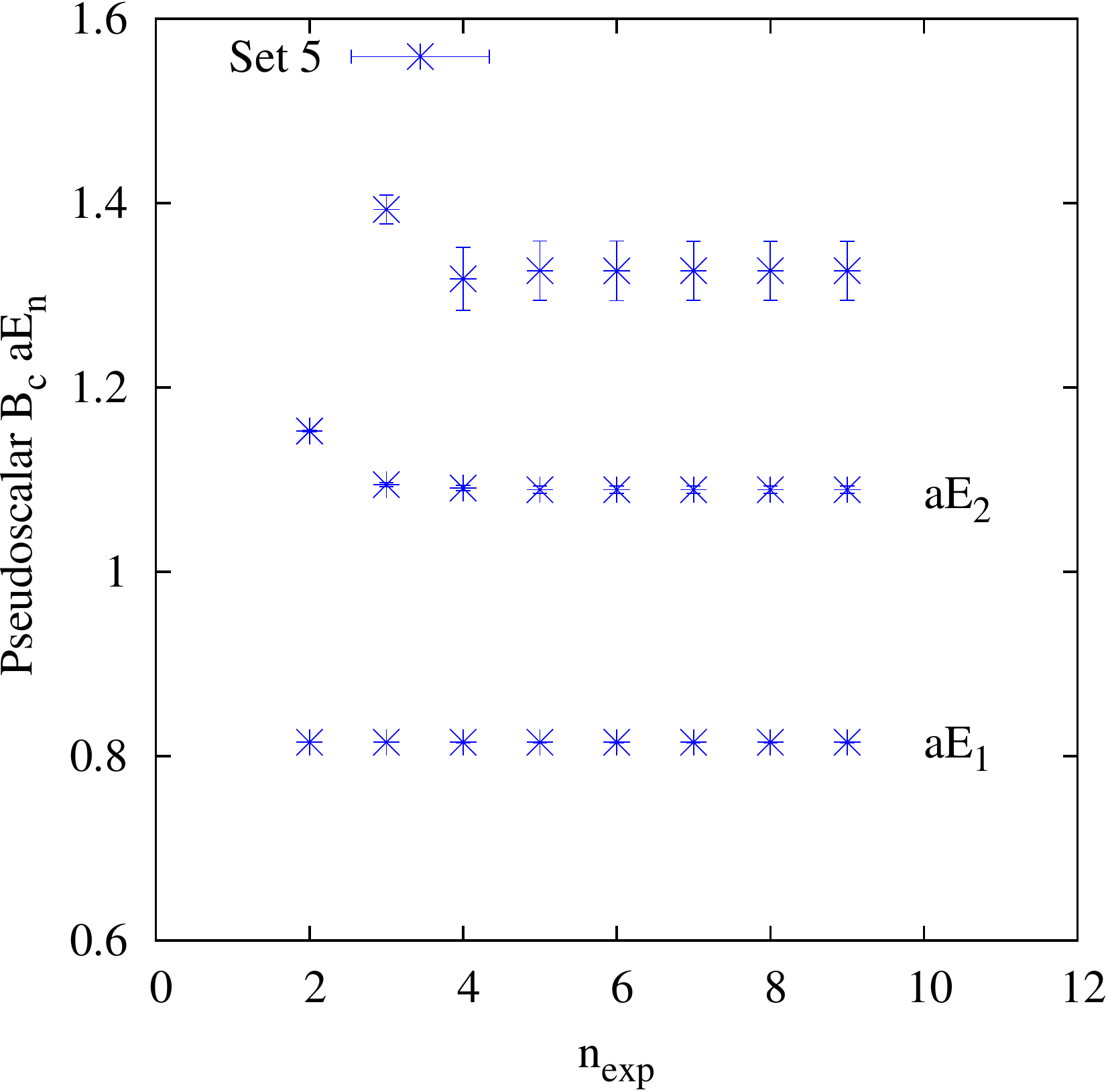}
\caption{
Energies of the 0$^-$ state in the $B_c$ fit against the number of exponentials in the fit on the fine ensemble, set 5. 
This shows how the value and the error on the first radial 
excitation energy, $aE_2$, stabilize as the number of exponentials is increased. 
}
\label{fig:Bcstar2s1sexp}
\end{figure}

Our $B_c$ meson correlators fits are accurate enough, and include multiple 
smearings to improve projection on the ground state, 
that there is a good signal for the first radially excited 
states, the $B_c'$ and $B_c^{*'}$. Fig. \ref{fig:Bcstar2s1sexp} shows how our 
fit results for these states converge. 
Unlike the $B$ and the $B_s$ these states are well below the threshold for strong decay, in this case into $B,D$, so that the states can be extracted unambiguously from a lattice calculation involving only operators that overlap onto single hadron states. The splittings from the ground state are listed in Table \ref{tab:bcresults}. We only have a signal for the coarse and fine ensembles since the starting time in the very coarse fits was set too high to extract excited states reliably. 

The splittings between the first radial excitation and the ground state 
are fit to the same form as the hs method in Eq.~\ref{eq:fitbchsextrap}, with a prior on 
the physical value of 0.5(5). Radial splittings are typically very 
insensitive to quark masses so we do not apply any shifts for mistuning. 
Such a shift would be dwarfed by the large statistical errors on 
the splittings. The only significant systematic error comes from missing radiative corrections to the spin-dependent terms in the action.
Since the ground state and radially excited state will be affected by this error in a similar way we take half the error applied in Sec. \ref{subsubsec:Bc-hh}, giving 1.2 MeV.

\begin{figure}
\includegraphics[width=0.8\hsize]{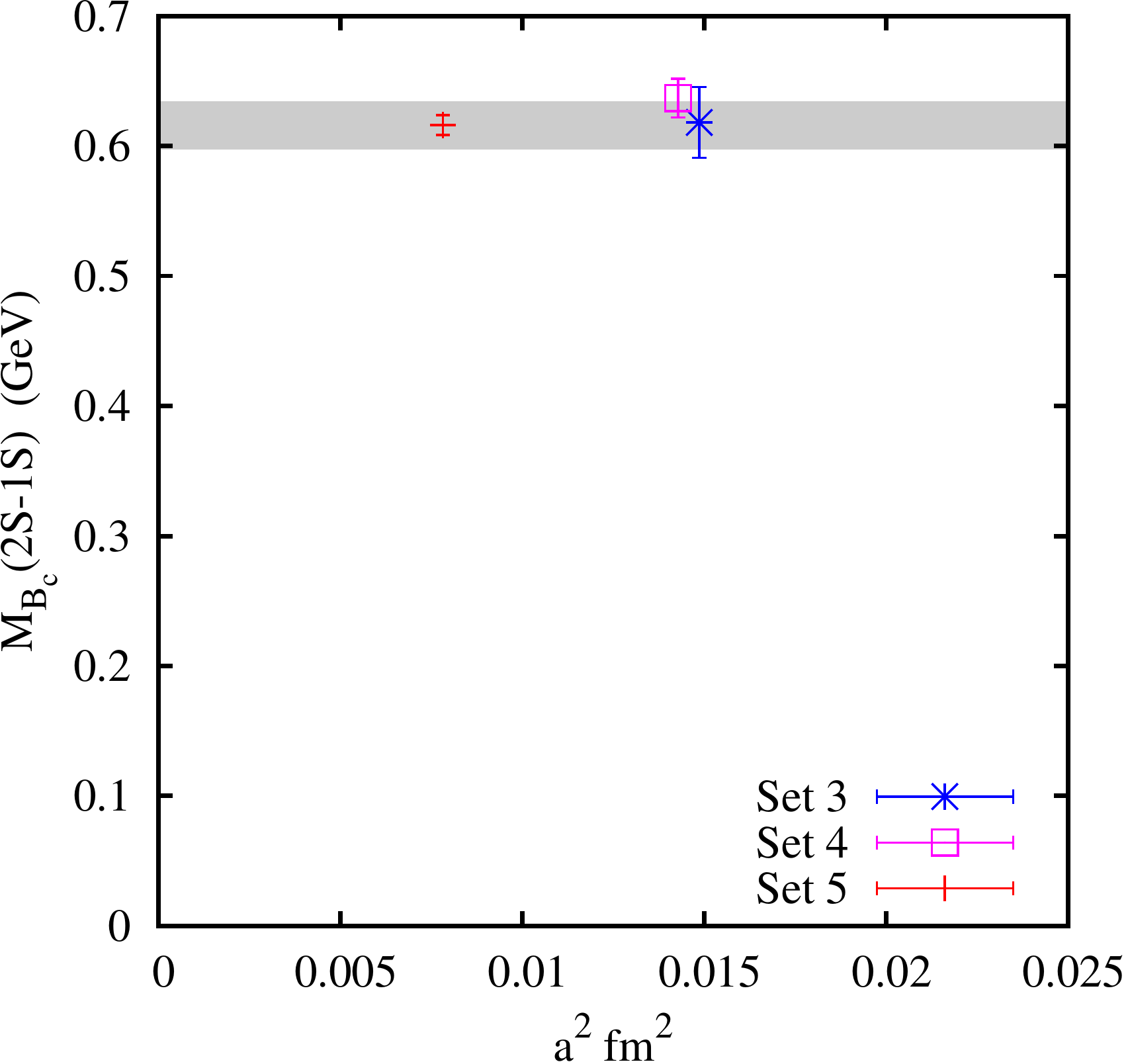}
\caption{
Results for the splitting $M_{B_c^\prime} - M_{B_c}$ on the fine and coarse ensembles along with the result of the fit.
}
\label{fig:Bc2s1s}
\end{figure}

\begin{figure}
\includegraphics[width=0.8\hsize]{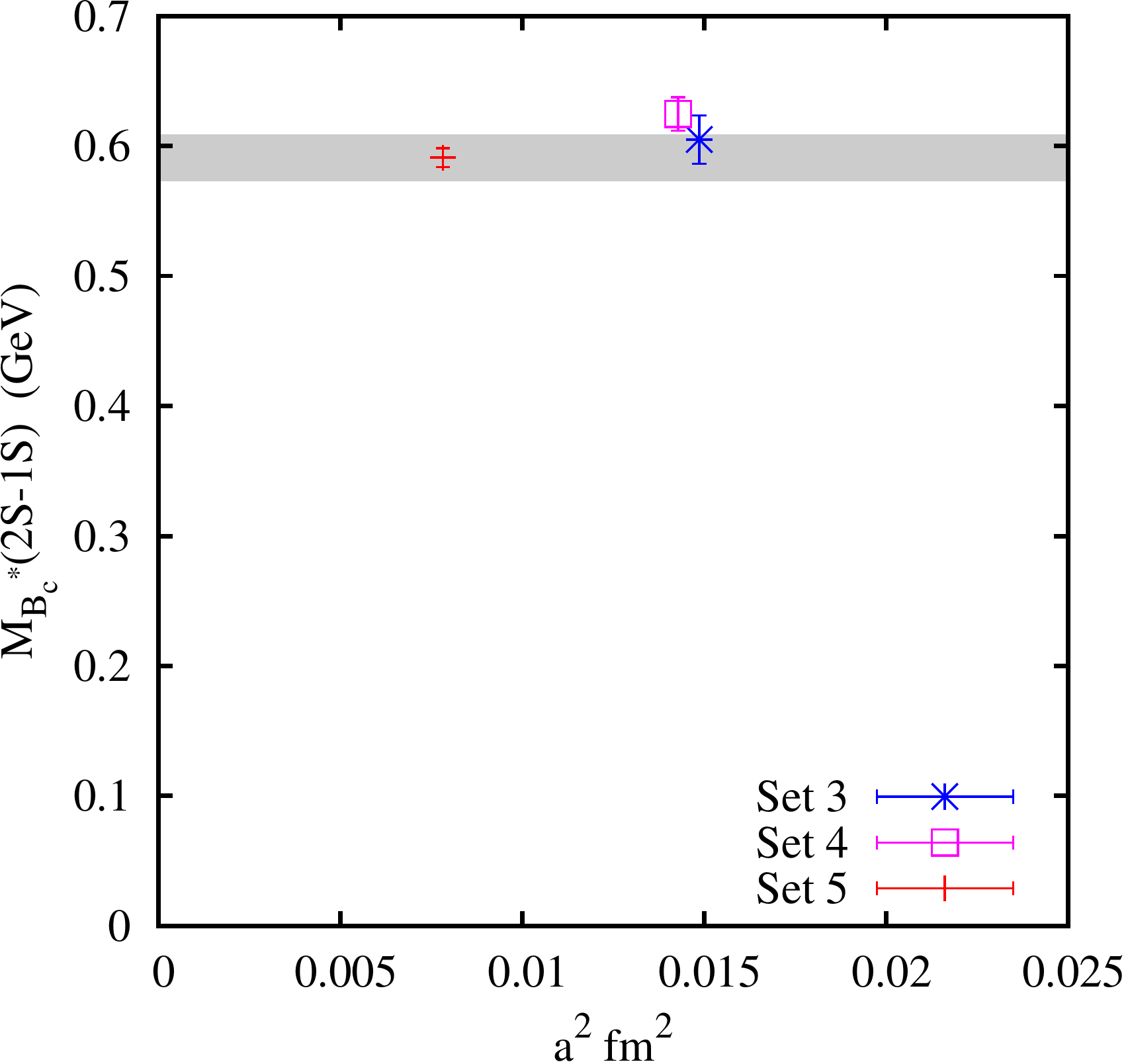}
\caption{
Results for the splitting $M_{B_c^{\ast\prime}} - M_{B_c^\ast}$ on the fine and coarse ensembles with the result of the fit.
}
\label{fig:Bcstar2s1s}
\end{figure}

The results from the fits are shown in Figs.~\ref{fig:Bc2s1s} and \ref{fig:Bcstar2s1s}, our results are:
\begin{eqnarray}
M_{B_c'} - M_{B_c}  &=& 616(19)\stat(1)\syst {\rm \ MeV}	\nn \\ 
M_{B_c^{*'}} - M_{B_c^*} &=& 591(18)_{\mbox{\tiny stat}} (1)_{\mbox{\tiny syst}} {\rm \ MeV},
\end{eqnarray}
where the error comes almost entirely from statistics/fitting.
The size of these splittings means that we expect the mesons to be sufficiently below threshold for strong decay into a $BD$ pair to be treated as gold-plated.
We are unable to resolve the excited hyperfine splitting.

The radial excitation energies for the $B_c$ can be compared to 
those for $\eta_c$ and $\eta_b$. For the $\eta_b$ recent Belle 
results~\cite{belleetab} give 0.597 GeV and for the $\eta_c$ the experimental 
average is 0.658 GeV~\cite{pdg}. Our $B_c$ result is between these two, 
as might be expected. 
For the $\Upsilon$ the experimental $2S-1S$ splitting 
is 0.563 GeV and for the $J/\psi$, 0.589 GeV~\cite{pdg}.
Our $B_c^*$ result agrees reasonably with either of these.  

%%%%%%%%%%%%%%%%%%%%%%%%%%%%%%%%%%%%%%%%%%%%%%%%%%%%%%%%%%%%%%%%%
%
\subsection{The $B$ meson}
\label{subsec:Bmeson}
%
%%%%%%%%%%%%%%%%%%%%%%%%%%%%%%%%%%%%%%%%%%%%%%%%%%%%%%%%%%%%%%%%%
\begin{table*}[ht]
\begin{ruledtabular}
\begin{tabular}{lllllllll}
Set & $am_b$ & $am_s$ & $am_l$ &  $aM_{\pi}$ & $aE(B_s)-aE(B)$ & $a\Delta^{\rm hyp}_{B}$ & $a\Delta^{0^+-0^-}_{B}$ &  $a\Delta^{1^+-1^-}_{B}$\\
\hline
1 & 3.297 & 0.0641 & 0.013 & 0.23637(15) & 0.05111(126) & 0.0375(12) & 0.245(17) & 0.251(20)\\ 
2 & 3.263 & 0.0636 & 0.0064 & 0.16615(7) & 0.05821(110) & 0.0377(9) & 0.207(25) & 0.150(57)\\ 
3 & 2.66 & 0.0522 & 0.01044 & 0.19153(9) & 0.04288(63) & 0.0324(4) & 0.193(13) & 0.192(15)\\ 
4 & 2.62 & 0.0505 & 0.00507 & 0.13413(5) & 0.04705(60) & 0.0309(4) & 0.200(4) & 0.207(4)\\ 
5 & 1.91 & 0.0364 & 0.0074 & 0.14070(9) & 0.03134(78) & 0.0212(11) & 0.159(8) & 0.158(7)\\ 
\end{tabular}
\end{ruledtabular}
\caption{
\label{tab:bresults}
Results in lattice units needed to determine the B meson mass. The first three columns give the $b,s$ and $l$ valence quark masses used in the runs. $aM_\pi$ is the pion mass calculated in \cite{Dowdall:2011wh} to be used in the chiral fits.
$aE(B_s)-aE(B_l)$ is the splitting between the $B_s$ and $B$ and $a\Delta^{\rm hyp}_{B}$ is the $B$ hyperfine splitting.
The final two columns give the splittings with the parity partner states discussed in Sec. \ref{sec:axialmesons}.
}
\end{table*}

We extract the mass of the $B$ meson using the 
splitting $\Delta_{B} =  M_{B_s} - M_{B}$ in which NRQCD systematics should cancel. 
The mass of the $B$ can then be reconstructed using our 
determination of $M_{B_s}$  in Sec. \ref{subsec:Bsmeson}. 
Results for the lattice energy splittings $aE_{B_s} - aE_{B}$ are 
given in Table \ref{tab:bresults} along with the values of $M_{\pi}$
on each ensemble needed for extrapolation in the light quark mass. 
The large correlation matrix meant that the correlators for 
each meson had to be fit separately but the statistical 
errors are a significant improvement over those in \cite{Gregory:2010gm}.

Heavy meson chiral perturbation theory (HM$\chi$PT) is used for the chiral fits. We use the 1-loop formulas given by Jenkins in \cite{Jenkins:1992hx} including heavy quark spin symmetry breaking terms at order $1/m_Q$, and up to $\mathcal{O}(M^3)$ in the light mesons masses.
Using the same notation as \cite{Jenkins:1992hx}, the full $SU(3)$ formula is
\begin{eqnarray}
 M_{B_s} - M_{B_d} &=& 
-\frac{3}{4}(2a +2\Delta^{(\sigma)})(m_s - m_l)
 \\ &&
-\frac{g^2 \pi}{\Lambda_{\chi}^2} \left[ \frac{3}{2}M_\pi^3 -2M_K^3 -\frac{1}{2} M_\eta^3   \right] \nn \\
&&+\frac{3g^2\Delta }{4\Lambda_{\chi}^2} \left[ -\frac{3}{2}l(M^2_\pi) + l(M^2_K) +\frac{1}{2}l(M_\eta^2)   \right]  \nn
\end{eqnarray}
where $\Lambda_{\chi} = 4\pi f_\pi$ is the chiral scale, $a$ and $\Delta^{(\sigma)})$ are coefficients of the tree level terms, $g$ is the $BB^\ast\pi$ coupling and $\Delta$ is the coefficient of the term in the effective Lagrangian that gives rise to the heavy meson hyperfine splitting. The chiral logarithms are given by
\begin{equation}
l(M^2) = M^2\left(  \ln \frac{M^2}{\Lambda^2} + \delta^{FV}(ML)      \right) 
\end{equation}
including the finite volume correction \cite{Bernard:2001yj}
\begin{equation}
\delta^{FV}(ML)  = \frac{4}{ML} \sum_{\vec{n} \neq 0 }  \frac{ K_1( |\vec{n}| ML ) } {|\vec{n}| }, 
\end{equation}
where $K_1$ is a modified Bessel function and the sum is over spatial vectors with components $n_i\in \mathbb{Z}$. 
The finite volume corrections shift the pion chiral logarithms by a few percent on some ensembles but have a completely negligible effect on the fit result.
We use the kaon and pion masses calculated in \cite{Dowdall:2011wh}, and use the tree level relation to change $M_\eta^2$ into a combination of $M_K^2$ and $M_{\pi}^2$. 
Quark masses are converted to meson masses using tree level relations.

Our central result uses the reduced $SU(2)$ version of the formula: 
\begin{eqnarray}
\label{eq:SU2HMChiPT}
 M_{B_s} - M_{B_d} &=& 
C -\frac{3}{4}(2a +2\Delta^{(\sigma)})m_l 
-\frac{g^2\pi}{\Lambda^2} \left[ \frac{3}{2}M_\pi^3   \right] \nn \\
&&+\frac{3g^2\Delta }{4\Lambda^2} \left[ -\frac{3}{2}l(M^2_\pi)   \right] 
\end{eqnarray}
for some constant $C$.
We also perform the fits using the $SU(3)$ formula as a check of systematic errors. 
Since we have a single pion mass for each ensemble and the sea strange quark masses are well tuned, partial quenching will be a small effect and we use only the full QCD form. Staggered quark and other discretisation effects could be more significant, however, so the fit function is multiplied by
\begin{equation}
 (1.0 + d_1(\Lambda a)^2 + d_2(\Lambda a)^4 )
\end{equation}
at a scale of $\Lambda=0.4$ GeV.

We take the prior on $g$ to be 0.5(5) which based on several recent lattice calculations \cite{Detmold:2012ge,Bulava:2010ej,Negishi:2006sc,Becirevic:2009yb} with a wide error covering all of the central values. 
Our results are not sufficient to constrain $g$, so we test the dependence of the final answer on this prior by varying its width. While this affects the shape of the curve, the result at the physical point does not change significantly since we have sufficiently light pion masses.
The prior on the tree level quark mass term is taken to be 0.5(5) and the priors on the discretisation terms $d_1,d_2$ are 0.0(5) and 0(1) respectively.

\begin{figure}
\includegraphics[width=0.9\hsize]{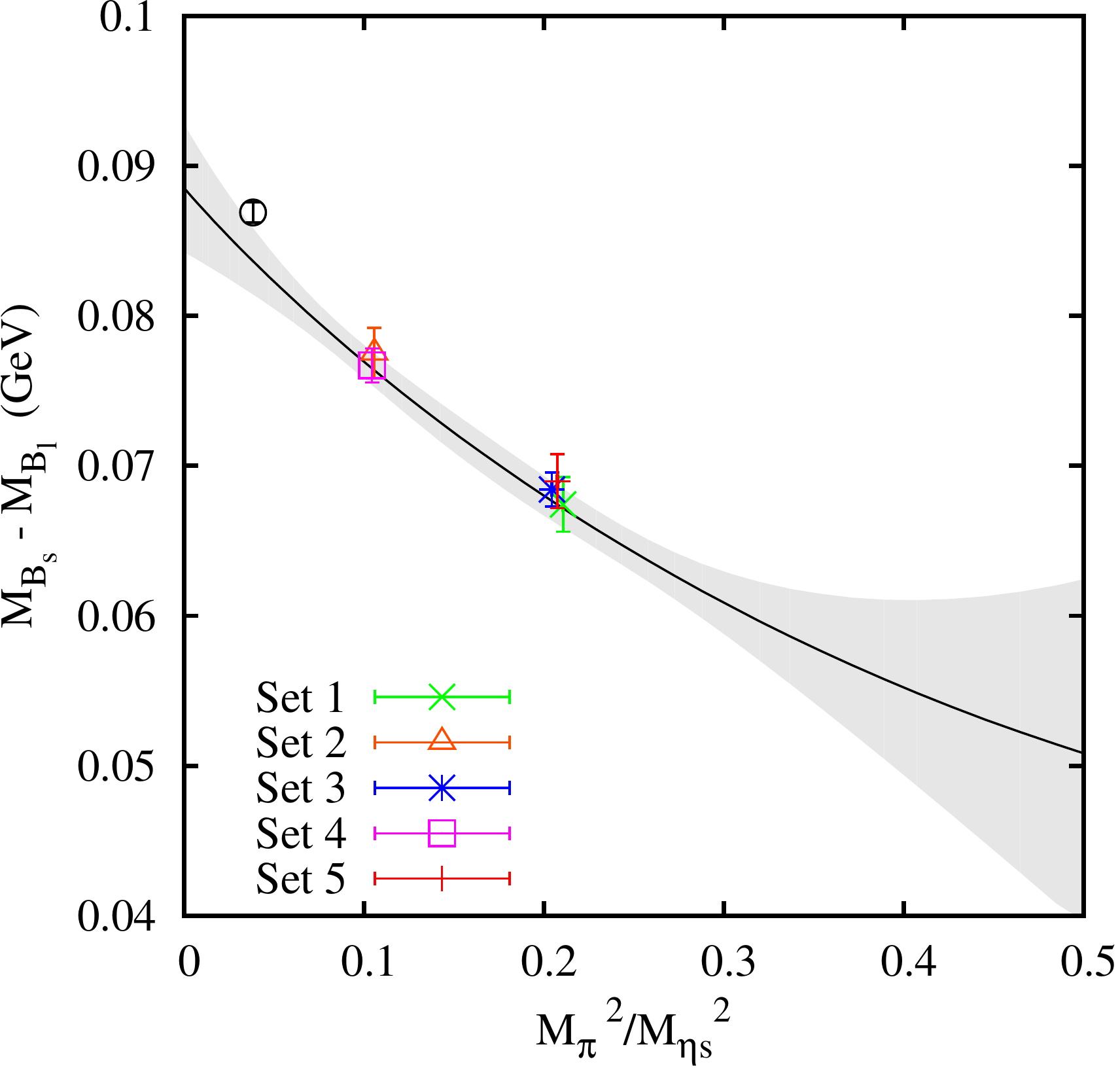}
\caption{
Chiral fit for $ M_{B_s} - M_{B_l}$ against the ratio $M_\pi^2 / M_{\eta_s}^2$ for the $SU(2)$ fit function with discretisation effects. The grey band shows the chiral fit evaluated at $a=0$ with no other systematic errors.
The plot includes the shift due to electromagnetic effects missing in lattice QCD.
}
\label{fig:Bd}
\end{figure}

\begin{figure}
\includegraphics[width=0.75\hsize]{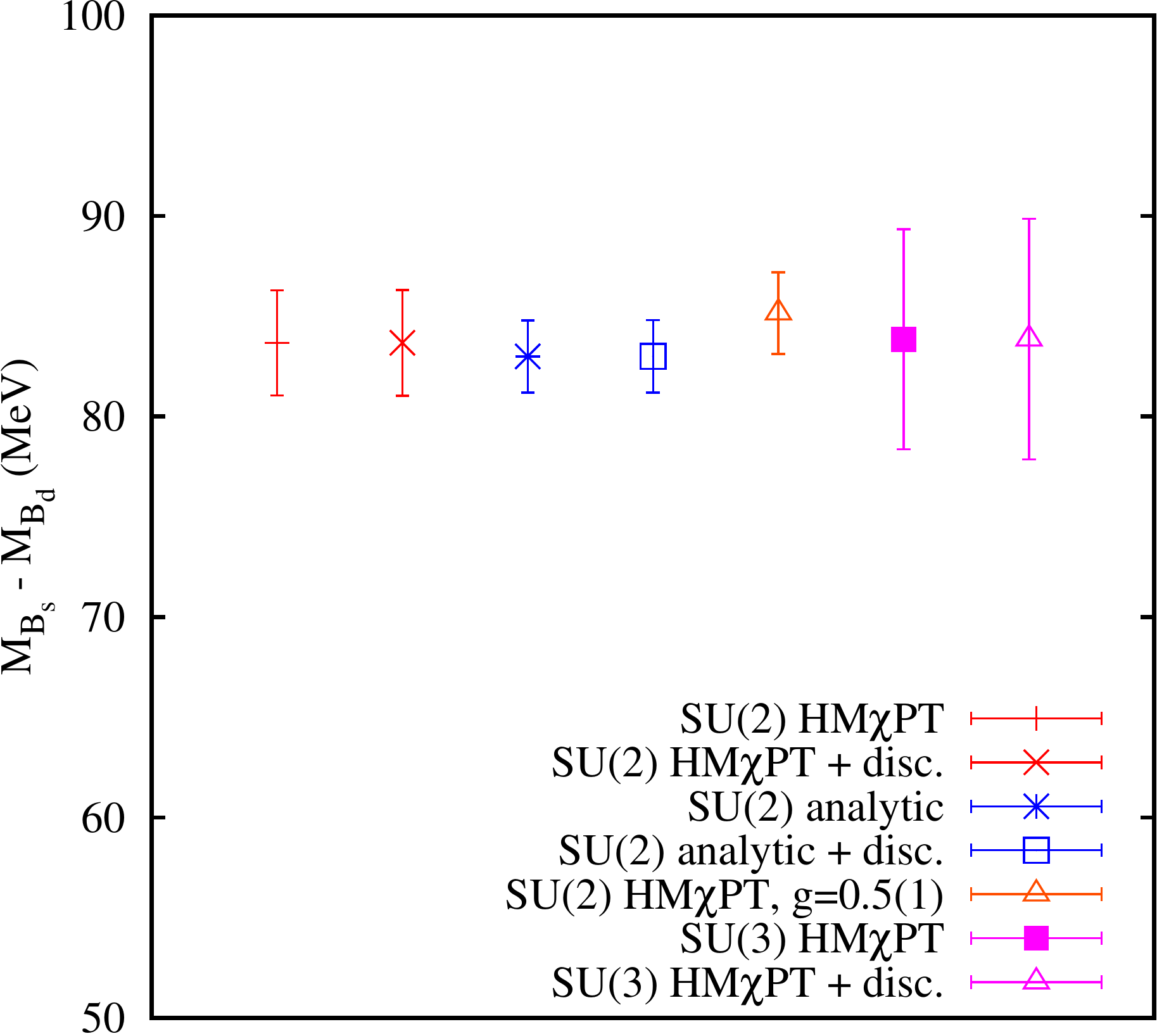}
\caption{
Comparison of different chiral fit functions: SU(2) HM$\chi$PT with and without discretisation corrections; leading order analytic terms with and without discretisation terms; SU(2) HM$\chi$PT with a tighter prior on $g$ of 0.5(1); SU(3) HM$\chi$PT with and without discretisation terms.
Only the error from the chiral fit is shown. 
}
\label{fig:Bd-comparison}
\end{figure}

The result of the $SU(2)$ fit is $ M_{B_s} - M_{B_d} = 85(2)$ MeV when 
evaluated at $a = 0$ and at the physical mass of the $\pi^0$ meson of 0.135 GeV. 
The fit is shown in Fig. \ref{fig:Bd} and gives a result around 1$\sigma$ below experiment. 
To check the reliability of the fit, the results of several different fit functions are plotted in Fig. \ref{fig:Bd-comparison}. This includes the 1-loop $SU(2)$ case, $SU(2)$ with different prior widths on $g$, the $SU(3)$ case and just the tree level terms with discretisation effects added in each case. Good $\chi^2$ values and consistent results are obtained for all fits.

We now need to consider the effect of electromagnetism on  $\Delta_{B_{\rm phys} } $. 
Since our light quark masses are degenerate we do not distinguish 
between the $B_d$ and $B_u$ mesons but compare to the 
average $M_{B_l} = (M_{B^\pm} + M_{B_0})/2$. Electromagnetism will 
affect the two states differently since the $B_u$ is charged. 
In \cite{Gregory:2010gm} the shift was estimated to be +2(1) MeV for 
the $B_u$ whereas the shift was negligible for the $B_d$ and $B_s$. 
So to compare with experiment we shift $\Delta_{B_{\rm phys} }$ by -1 MeV to give 
$$
M_{B_s} - M_{B_l} = 84(2) {\rm \ MeV}
$$
in good agreement with experiment of 87.4(3) MeV (within 2 sigma).
Reconstructing $M_B$ using our value for $M_{B_s}$ in Sec. \ref{subsec:Bsmeson} gives
$
M_B = 5.283(2)(8) {\ \rm GeV}
$
The first error is from the chiral fit and the second is 
the error on $M_{B_s}$ with the detailed breakdown as in Sec. \ref{subsec:Bsmeson}.

%%%%%%%%%%%%%%%%%%%%%%%%%%%%%%%%%%%%%%%%%%%%%%%%%%%%%%%%%%%%%%%%%
%
\subsection{The $D$ meson}
\label{subsec:Dmeson}
%
%%%%%%%%%%%%%%%%%%%%%%%%%%%%%%%%%%%%%%%%%%%%%%%%%%%%%%%%%%%%%%%%%
\begin{figure}
\includegraphics[width=0.9\hsize]{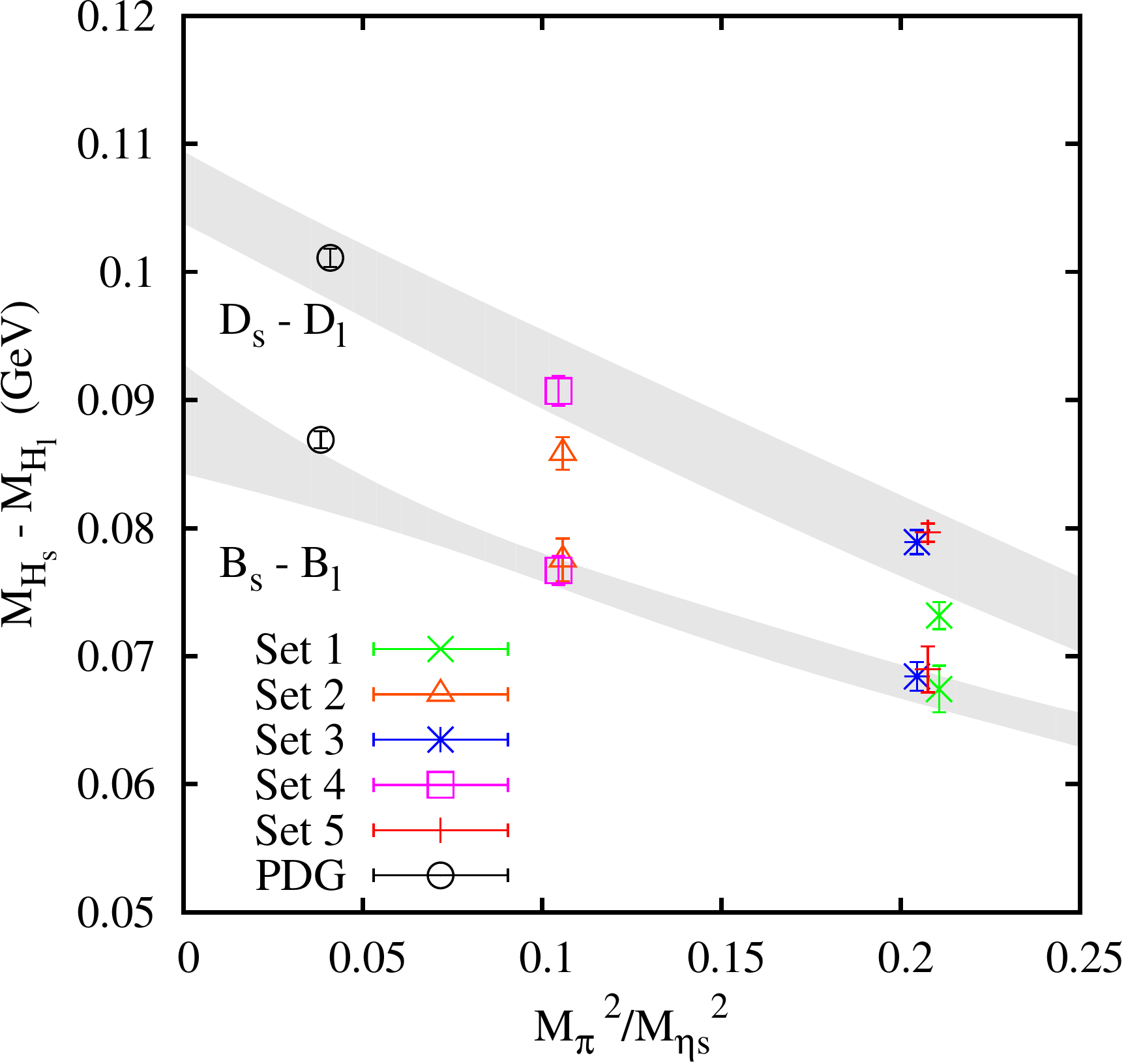}
\caption{
Plot of the chiral extrapolation for $M_{D_s} - M_{D}$ with $M_{B_s} - M_{B}$ for comparision.
Errors are from the chiral fit only and the lattice data is adjusted for missing electromagnetism.
}
\label{fig:D}
\end{figure}

Our analysis of the $D$ meson follows the same method as the $B$ in the previous section. 
The splitting $M_{D_s}-M_D$ is taken from a combined fit to all three 
charmed mesons, the results are given in Table~\ref{tab:Dresults}. 
Systematic errors should be small in the splitting since the only 
difference between the states is the light quark mass, 
however we still see some lattice spacing dependence coming 
from the charm quark discretisation errors. 
We use the SU(2) HM$\chi$PT formula Eq.~\ref{eq:SU2HMChiPT} with 
discretisation terms, this time including higher powers of $a$ 
and with a scale set by $m_c$
\begin{equation}
 (1.0 + d_1(m_c a)^2 + d_2(m_c a)^4 + d_3(m_c a)^6 + d_4(m_c a)^8).
\end{equation}
Priors for $g$ and the leading term are the same as above but priors 
for discretisation errors are 0.00(6) for $d_1$ and 0.0(2) for other $d_i$ 
terms as in Sec.~\ref{subsec:Dsmeson}. 
Since we do not have vector meson masses
the experimental value 140 MeV~\cite{pdg} is used for the 
hyperfine term in the fit function. The HISQ action has previously 
been shown to give results for hyperfine splittings in agreement 
with experiment~\cite{Follana:2006rc}. 

The result of the fit is shown in Fig.~\ref{fig:D} including an adjustment for electromagnetism. 
The shift in the $D_s$ is 1.3(7) MeV and the shifts in the 
$D_0$ and $D^\pm$ are -0.4 MeV and +1.3 MeV, which results in a total 
shift of 0.9 MeV in $M_{D_l} = (M_{D_0} + M_{D^\pm})/2$.
As for the $B$, tightening the prior on $g$ to 0.1 also gives a consistent result, but in this case discretisation errors are significant so removing the $d_i$ terms leads to a poorer fit.

Our final result for the splitting is
$$
M_{D_s} - M_{D_l} = 101(3) {\rm \ MeV}
$$
in agreement with the experimental splitting of 101.3(3) MeV~\cite{pdg}. 
When combined with our result for $M_{D_s}$ above this 
gives $M_{D_l} = 1.869(3)(3) \mathrm{GeV}$, 
the first error being the chiral fitting error and the second the full error from $M_{D_s}$. 

Fig.~\ref{fig:D} shows $M_{B_s} - M_B$ and $M_{D_s}-M_D$ on the 
same plot.  It is clear that lattice QCD can distinguish the difference 
between these two small splittings. In HQET language it arises from 
the difference in the kinetic energy of the heavy quark in a heavy-strange 
meson compared to that in a heavy-light meson. We would expect this 
difference to be positive and contribute a larger amount for $c$ 
quarks than $b$ quarks, consistent with the increase seen. 
It is clear that lattice QCD successfully reproduces this effect. 

%%%%%%%%%%%%%%%%%%%%%%%%%%%%%%%%%%%%%%%%%%%%%%%%%%%%%%%%%%%%%%%%%
%
\section{Hyperfine splittings}
\label{sec:hyperfines}
%
%%%%%%%%%%%%%%%%%%%%%%%%%%%%%%%%%%%%%%%%%%%%%%%%%%%%%%%%%%%%%%%%%

\begin{table*}
\begin{ruledtabular}
\begin{tabular}{llllll}
Set& $a\Delta_{B}^{\mbox{\tiny hyp}}$ &$a\Delta_{B_s}^{\mbox{\tiny hyp}}$ & $a\Delta_{B_c}^{\mbox{\tiny hyp}}$ 
& $m_b^{\rm phys}$ & tuning \\
\hline
1 & 0.0375(12) & 0.03892(40) & 0.03659(17) & 3.297(11)(35)(7)(16) & 1.000(5)\\ 
2 & 0.0377(9) & 0.03705(47) & 0.03658(13) & 3.263(7)(35)(4)(16) & 1.000(5)\\ 
3 & 0.0324(4) & 0.03177(18) & 0.03140(3) & 2.696(4)(22)(7)(13) & 0.987(5)\\ 
4 & 0.0309(4) & 0.03102(16) & 0.03137(2) & 2.623(7)(22)(7)(13) & 0.999(6)\\ 
5 & 0.0212(11) & 0.02310(14) & 0.02470(2) & 1.893(6)(12)(5)(9) & 1.009(5)\\ 
\end{tabular}
\end{ruledtabular}
\caption{\label{tab:hypresults}
Results for the hyperfine splittings $\Delta_{B_q}^{\mbox{\tiny hyp}}$ in lattice units for each ensemble, errors are statistical only. Column 5 gives the tuned $b$ quark masses calculated in \cite{Dowdall:2011wh} where the first two errors are from statistical and systematic errors respectively in the lattice spacing determination. The third and fourth
errors are the statistical and systematic errors
in the determining the Upsilon kinetic mass used for tuning $am_b$.
The final column gives the multiplicative factor applied to each hyperfine splitting due to $b$ quark mass mistuning. 
}
\end{table*}
\begin{figure*}
\includegraphics[width=0.32\hsize]{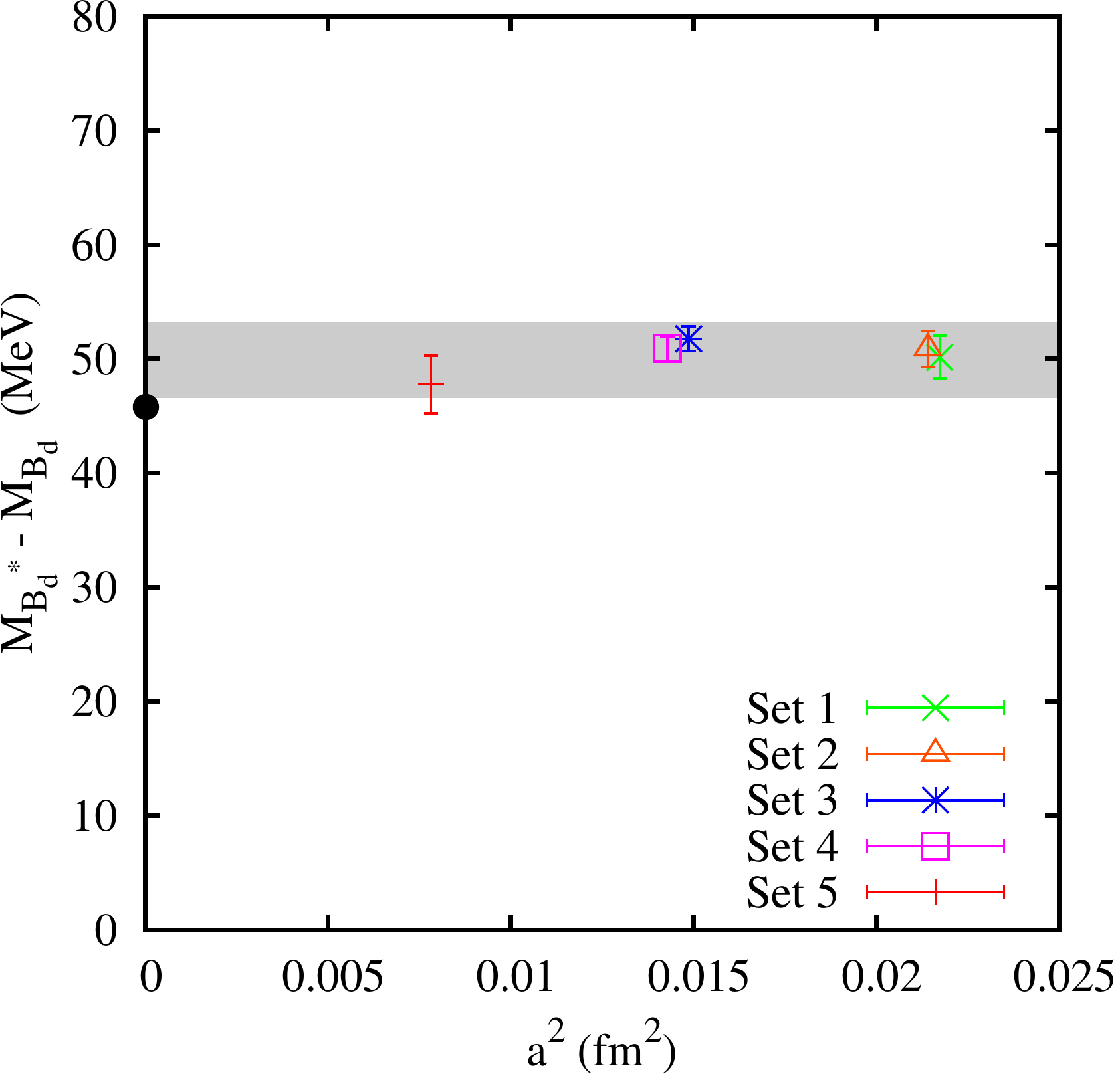}
\includegraphics[width=0.32\hsize]{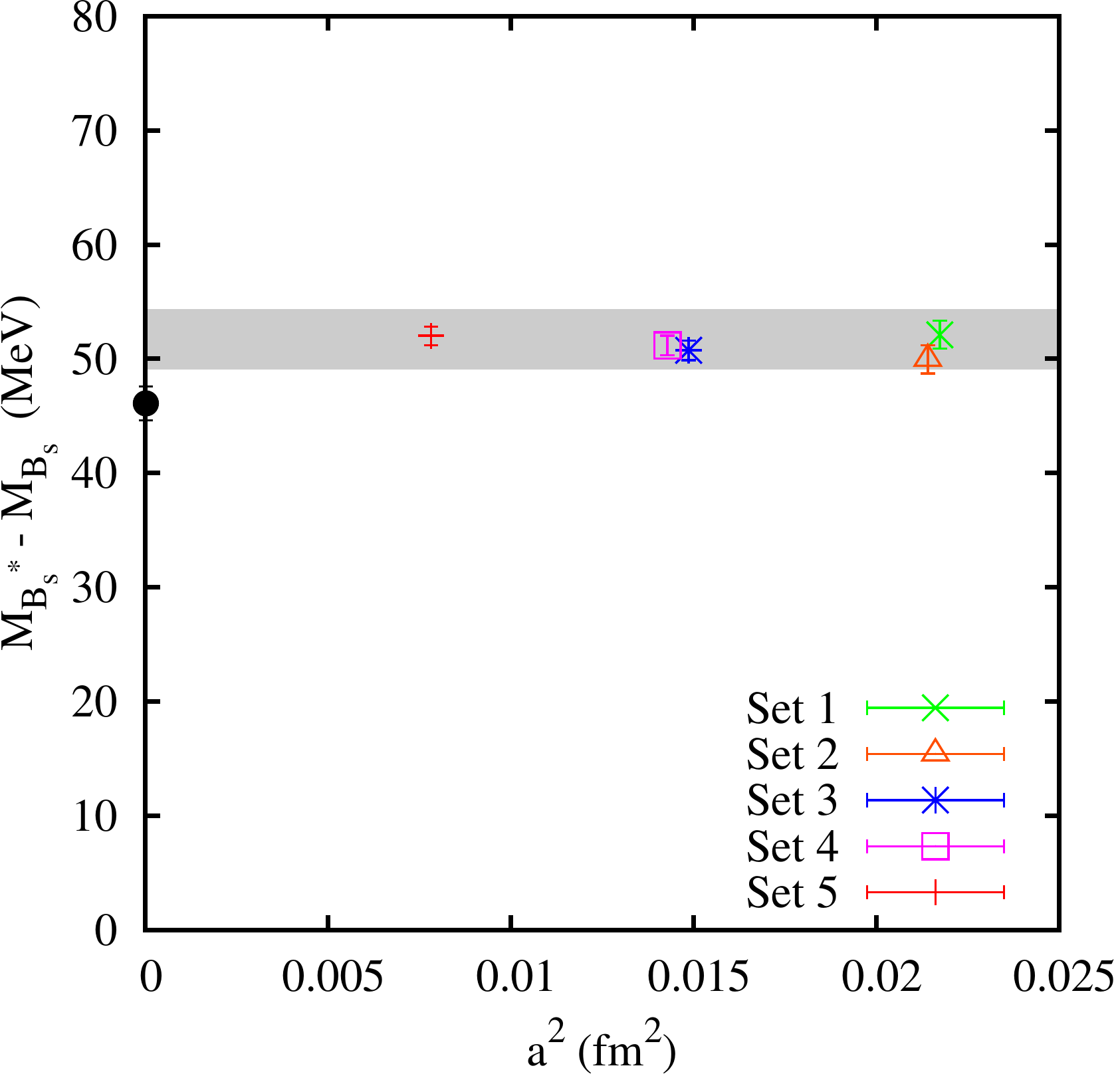}
\includegraphics[width=0.32\hsize]{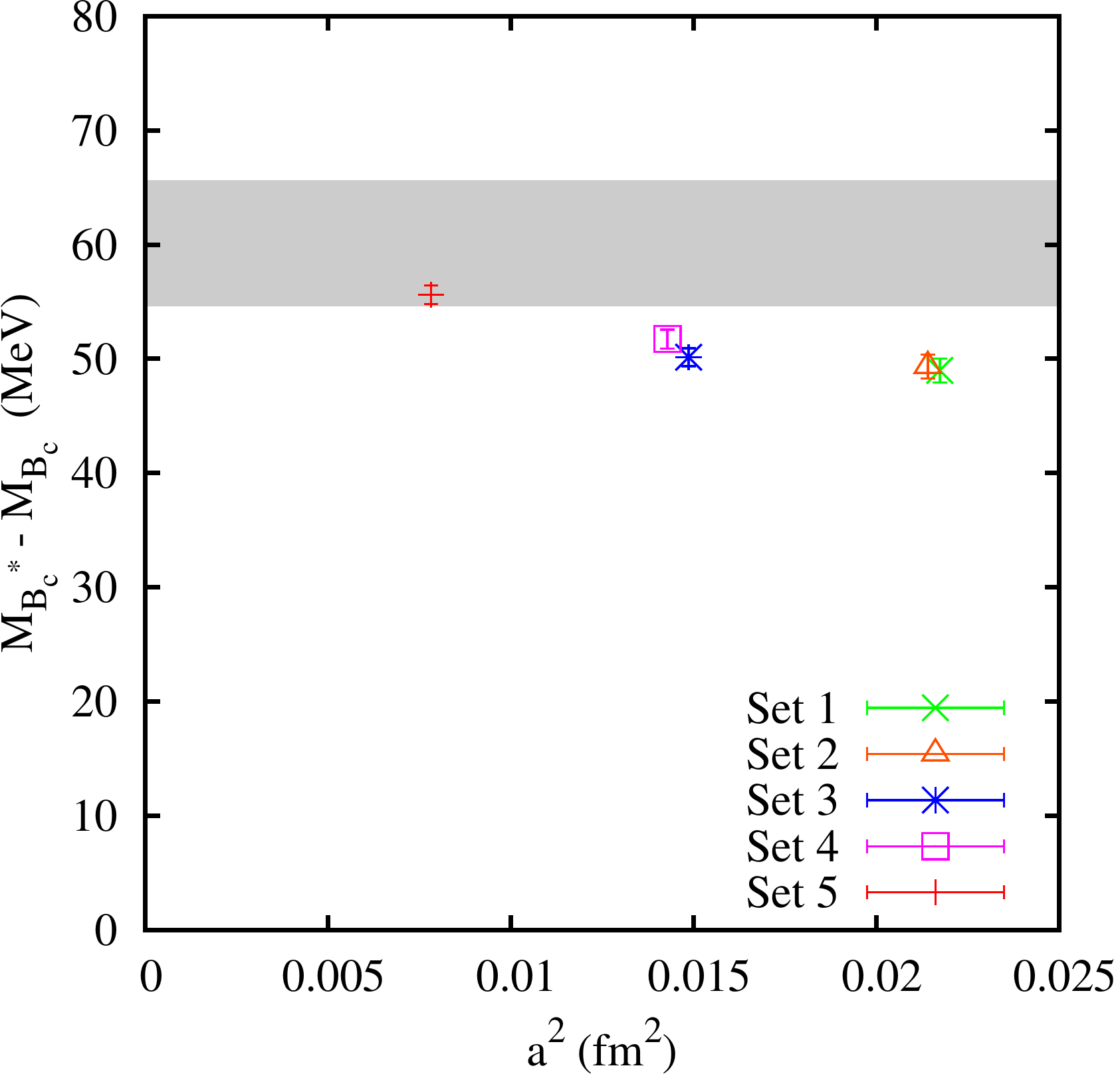}
\caption{ Results for the $B, B_s$ and $B_c$ hyperfine splittings. The data points include statistical, tuning and lattice spacing errors, taken as double the naive error as discussed in the text. The correlated $\alpha_s^2$ error is not included on the data points. 
The result of the fit in each case is shown as a grey band and, where available, experimental values are given as black solid circles.
}
\label{fig:hyp}
\end{figure*}
The hyperfine splitting between the ground state vector and pseudo-scalar 
states is a particularly good test of a spectrum calculation.
For heavy-light mesons this splitting is proportional to the term
$
\frac{c_4}{am_b} \sigma \cdot B
$
in the NRQCD action so gives a direct check of the radiative corrections to $c_4$. 
This is in contrast to the case in heavyonium where the hyperfine splitting 
is proportional to $c_4^2$. 
The splitting also depends on higher order operators but in heavy-light systems these terms will be very small, unlike in bottomonium where the $v^6$ terms could be 10\%.
The splitting $\Delta_{B_q}^{\mbox{\tiny hyp}} = M_{B_q^*} - M_{B_q}$ is very precise and including both the vector and pseudo-scalars in the same fit takes account of the correlations between the two.

The bottomonium hyperfine splittings were calculated using our improved action in \cite{Dowdall:2011wh}, where we obtained 70(9) MeV for the 1S hyperfine splitting and 0.499(42) for the ratio of the 2S and 1S hyperfine splittings (which agreed well with 
subsequent experiment~\cite{belleetab}). 
The error in both cases was dominated by the missing $v^6$ terms. 
The heavy-light hyperfine splittings have previously been studied in 
Ref.~\cite{Gregory:2009hq} by considering ratios that were independent of $c_4$. 
This resulted in a prediction of 53(7) MeV for the $B_c$ hyperfine splitting. 
The advantage of our current calculation is that the coefficients have been obtained by matching NRQCD to QCD at one loop, allowing the hyperfine splittings to be determined directly without losing predictive power. Using the same action and $c_4$ for both the bottomonium and B-meson calculations also allow us to make very different, independent checks.

The results for the hyperfine splittings are given in lattice units in Table \ref{tab:hypresults} for the $B_l,B_s$ and $B_c$. 
Before fitting the data we make a small correction for the $b$ quark mass mistuning on each ensemble. The splittings are very insensitive to the light quark mass so retuning 
for $m_s,m_c$ will be negligible compared to other errors. 
The retuning assumes that the hyperfine splitting is inversely proportional 
to the $b$ quark mass and is applied multiplicatively using 
the tuned $b$ quark mass values $m_b^{\rm phys}$ calculated 
in \cite{Dowdall:2011wh} and listed in Table \ref{tab:hypresults}. There are two sources of error in $m_b^{\rm phys}$, coming from the lattice spacing and the determination of the bottomonium kinetic mass values. Since a change in the lattice spacing would result in a change in the quark mass, the lattice spacing uncertainty is correlated with the scale uncertainty in the hyperfine splitting itself. To account for this correlation, we apply twice the lattice spacing error to the hyperfine splitting rather than adding them separately. The retuning 
factors are all less than 2\% and are given in Table \ref{tab:hypresults}.

The dominant source of uncertainty in $\Delta_{B_q}^{\mbox{\tiny hyp}}$ is still the 
higher order correction to $c_4$ which is now $\mathcal{O}(\alpha_s^2)$. 
To allow for this we apply a correlated systematic error to all 
the data points of size $\alpha_s^2$ where we take 
$\alpha_s$ at a scale $\pi/a$. Values for $\alpha_s$ are: 0.275 on very coarse, 
0.255 on coarse and 0.225 on fine \cite{Dowdall:2011wh}.
The $B_s$ and $B_l$ hyperfine splittings are fit to the same form 
as $\Delta_{B_s}$ (Eq.~(\ref{eq:fitbsextrap})), allowing for lattice 
spacing, sea quark mass and cutoff dependence. 
The $B_c$ hyperfine is fit to the form used for $\Delta_{B_c}$ in 
the hs method (Eq.~(\ref{eq:fitbcextrap})) in which we include discretisation 
errors with a scale set by $m_c$.
Priors are the same in all cases except for the prior on $\Delta_{B_q}^{\mbox{\tiny hyp}}$ which is 0.5(5).

The data and fit results are plotted in Fig. \ref{fig:hyp} with the data points adjusted for $b$ quark mass mistuning but not including the correlated systematic error from $c_4$. There is no noticable sea quark mass dependence and very little $a$ dependence except for the $B_c$ case where the discretisation errors come from the charm quark.
The dependence of $\Delta_{B_q}^{\mbox{\tiny hyp}}$ on the light valence quark is also very small and not statistically significant.
The results for the physical values from the fits are:
\begin{eqnarray}
 \Delta_{B_l}^{\mbox{\tiny hyp}} &=& 50(3)  {\rm MeV}
\nn \\
\Delta_{B_s}^{\mbox{\tiny hyp}} &=& 52(3)  {\rm MeV}
\nn \\
\Delta_{B_c}^{\mbox{\tiny hyp}} &=& 60(6) {\rm MeV}.
\end{eqnarray}

The other remaining source of error is the effect 
of $v^6$ terms which are very small here.
The full error budget for each splitting is given in Table \ref{tab:hyperr}. 
Comparison to experiment (45.8(4) MeV~\cite{pdg}) 
for $B_l$ shows good agreement. 
For $B_s$ the experimental results are not as accurate. 
In Fig.~\ref{fig:hyp} we use the experimental 
average of 46.1(1.5) MeV~\cite{pdg}. This agrees with 
our value within $2\sigma$.

\begin{table}[t]
\caption{The full error budget for the hyperfine splittings, giving each error 
as a percentage of the final answer. The fit value is obtained including the statistical, scale and $\alpha_s^2$ errors and their separate contribution to the error budget is distinguished by fitting with and without the $\alpha_s^2$ error.
$v^6$ errors are included multiplicatively using the estimates in Sec.~\ref{sec:lattice}.
The error from $a$, $m_{q,sea}$, and $am_b$ dependence is estimated from the fit.
The error from $m_b$ tuning is estimated by fitting with and without the error on the tuning in Table \ref{tab:hypresults}.
   }
\label{tab:hyperr}
\begin{ruledtabular}
\begin{tabular}{llllll}
 & $\Delta_{B}^{\mbox{\tiny hyp}}$ &$\Delta_{B_s}^{\mbox{\tiny hyp}}$ & $\Delta_{B_c}^{\mbox{\tiny hyp}}$ 
& $R_{B}$ & $R_{B_c}$\\
\hline
stats/fitting/scale	& 2.0	& 1.9	& 5.8   & 2.3 &	1.5 \\
$a$-dependence 		& 1.3	& 0.8	& 3.6   & 2.1 &	2.5 \\
$m_{q,sea}$-dependence 	& 1.6	& 1.7	& 2.8   & 1.0 &	1.2 \\
NRQCD $am_b$-dependence & 0.1	& 0.6	& 5.3   & 0.2 & 3.7 \\
NRQCD $v^6$ 		& 0.1	& 0.1	& 0.5   & 0.5 & 2.0 \\
NRQCD $c_4$ uncertainty & 6.0	& 4.4	& 4.7   & 0.0 & 0.0 \\
$m_b$ tuning	 	& $<$0.1&$<$0.1	&$<$0.1 & 0.0 &	0.0 \\
\hline
Total 	(\%)		& 6.7 	& 5.2	& 10    & 3.3 & 5.2    
\label{tab:hyperrors}
\end{tabular}
\end{ruledtabular}
\end{table}

\begin{figure*}
\includegraphics[width=0.4\hsize]{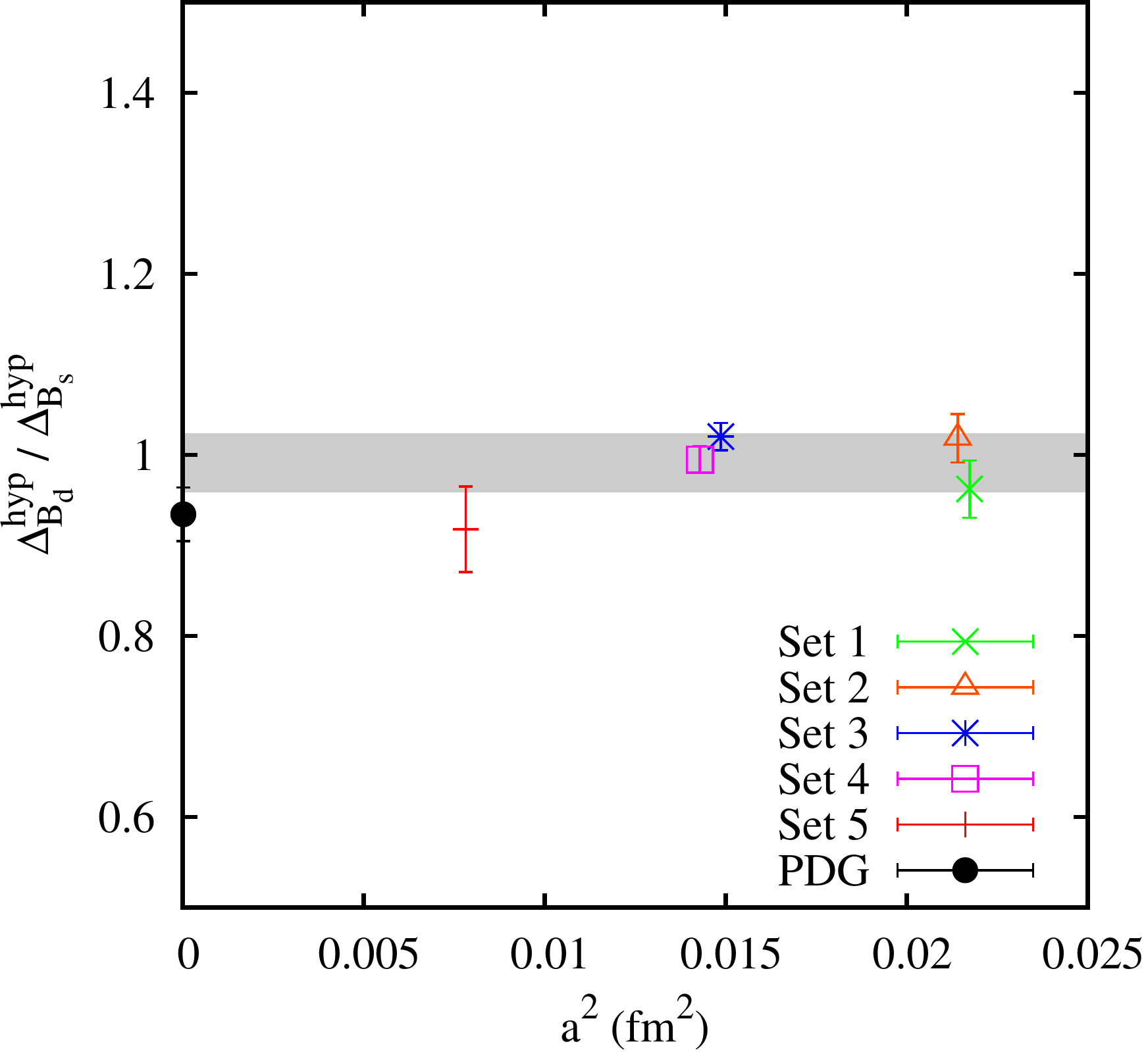}
\includegraphics[width=0.4\hsize]{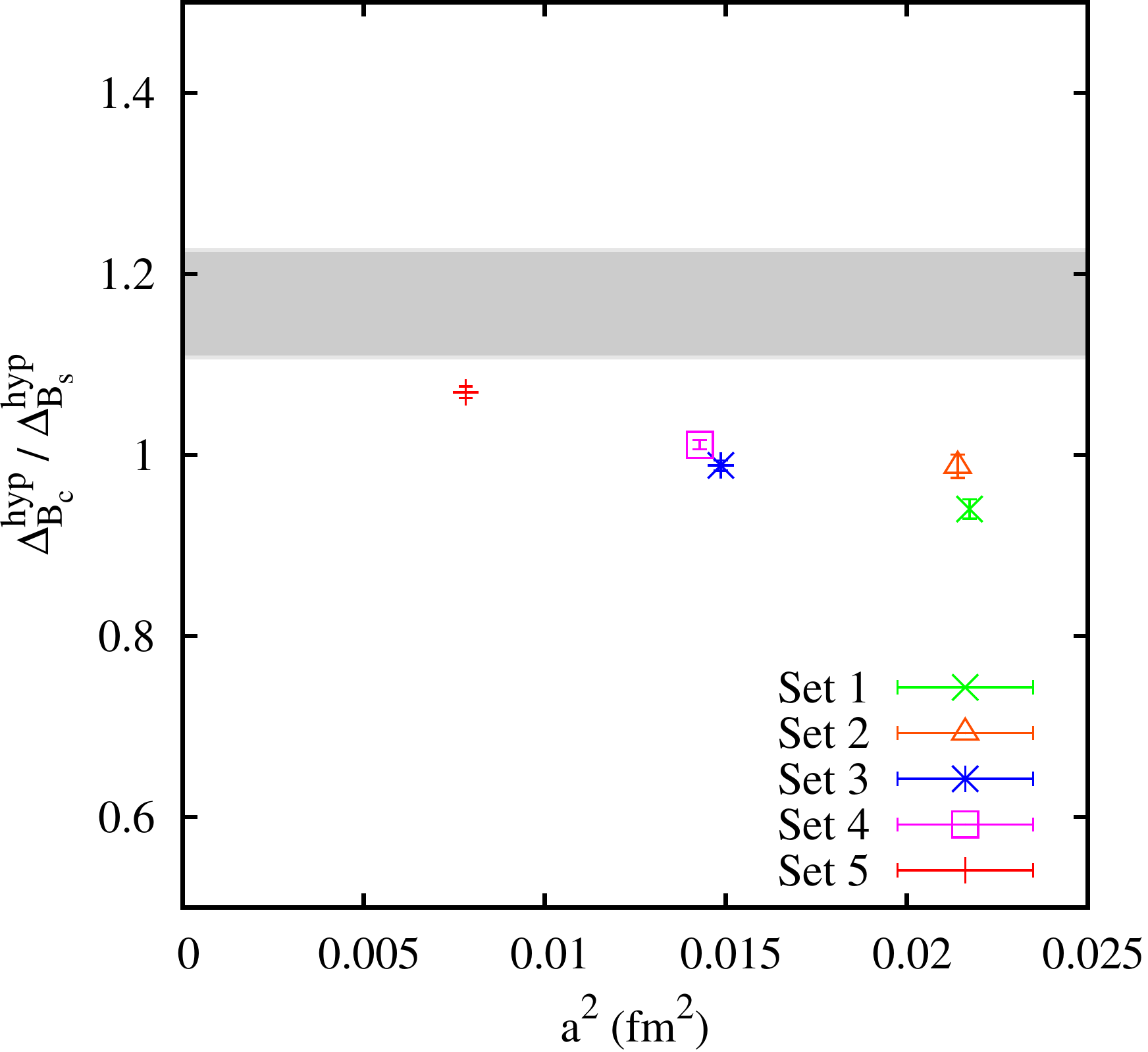}
\caption{ Ratios of B-meson hyperfine splittings.
The result of the fit in each case is shown as a grey band and, where available, experimental values are given as black solid circles.
}
\label{fig:hypratios}
\end{figure*}

The dominant error in the hyperfine splittings is still the uncertainty 
in the $c_4$ coefficient which is reduced in this calculation to $\mathcal{O}(\alpha_s^2)$. 
Taking ratios of hyperfine splittings eliminates this error and also cancels errors from the lattice spacing and mistuning of the $b$ quark mass. The remaining errors will be from missing $v^6$ terms which are very small.
Fig. \ref{fig:hypratios} shows results for the ratio of the $B$ and $B_c$ hyperfine splittings to that of the $B_s$. The 
fit function for ${ \Delta_{B_d}^{\rm hyp} }/{ \Delta_{B_s}^{\rm hyp} }$ is the same as for the $B_d$ hyperfine splitting and the fit function for ${ \Delta_{B_c}^{\rm hyp} }/{ \Delta_{B_s}^{\rm hyp} }$ is the same as for the $B_c$ hyperfine splitting. Priors on the ratios are taken to be 1.0(5).

The results of the fits are
\begin{eqnarray}
R_{B} = 
\frac{ \Delta_{B_l}^{\rm hyp} }{ \Delta_{B_s}^{\rm hyp} } &=&
0.993(33)(5) 
\nn \\
R_{B_c} = 
\frac{ \Delta_{B_c}^{\rm hyp} }{ \Delta_{B_s}^{\rm hyp} } &=&
1.166(56)(23).
\end{eqnarray}
The first error is from statistics/fitting and the second is the systematic error that is dominated by missing $v^6$ terms in the action. We take half the estimated size of $v^6$ terms as there should be some cancellation between the splittings.
The full error budget is given in Table \ref{tab:hyperrors}.
Our results are now precise enough that we are able to resolve the difference 
from 1.0 in the charm/strange hyperfine ratio, at the same time confirming 
our previous result~\cite{Gregory:2009hq}  
that this ratio is not far from 1. 
Our value for the ratio of light to strange hyperfine splittings is 1 with 
an accuracy of 3\% (equivalent to 1.5 MeV for this splitting). 
The experimental ratio, using the $B_s$ average above is 0.993(34). 

The accuracy of the ratios above means that we can give an improved 
prediction of $M_{B_c^*}-M_{B_c}$. Multiplying the experimental 
average for $B_s$ by the ratio above gives:  
\begin{equation}
\Delta_{B_c}^{\mbox{\tiny hyp}} = 54(3) {\rm MeV}.
\label{eq:bcfinal}
\end{equation}
We take this as our final predicted value. Note that this 
is smaller than either the bottomonium or charmonium ground-state 
hyperfine splittings. 

%%%%%%%%%%%%%%%%%%%%%%%%%%%%%%%%%%%%%%%%%%%%%%%%%%%%%%%%%%%%%%%%%
%
\section{Axial vector and scalar $B_c$ mesons}
\label{sec:axialmesons}
%
%%%%%%%%%%%%%%%%%%%%%%%%%%%%%%%%%%%%%%%%%%%%%%%%%%%%%%%%%%%%%%%%%

As discussed in Sec.~\ref{sec:lattice}, our $B$-meson correlators 
contain oscillating terms corresponding to states of opposite parity. 
Hence our pseudoscalar correlators contain 
both 0$^-$ and 0$^+$ states, and the vector correlators contain 1$^-$ and 1$^+$ states. 
By using the fit form given in Eq.~(\ref{eq:fitform}) we can then 
extract the energies of these scalar and axial vector states 
from our fits. 

The splittings 
\begin{eqnarray}
a\Delta_{B_q}^{0^+-0^-} &=& aE_{B_{q0}^*} - aE_{B_{q}} \\
a\Delta_{B_q}^{1^+-1^-}&=& aE_{B_{q1}} - aE_{B_{q}^*}
\end{eqnarray}
are given in Tables~\ref{tab:bsresults},~\ref{tab:bcresults} and~\ref{tab:bresults} 
for the three mesons with $q=s,c,l$ respectively. We find that, for the $B$, both 
states are above threshold for decay into $B\pi$ and for the $B_s$ the states 
are very close to threshold for $BK$ decay, as was found in~\cite{Gregory:2010gm}. 
Since we do not have enough data to accurately estimate threshold effects in these cases we do not analyse them further, but they are included for completeness.

The $B_c$ states, however, are far enough below threshold for decay to $BD$ that we can reliably predict their masses. 
The remaining problem comes from identifying which states our results correspond to. 
From heavy quark spin symmetry, the ``P-wave'' heavy-light mesons come in two doublets, a 0$^+$, 1$^+$ pair coming from a light quark spin of $j_l=1/2$ and a 1$^+$, $2^+$ pair from  $j_l=3/2.$ Identifying our scalar state with the physical 0$^+$ state is unambiguous but the situation is not as clear for the axial-vector. Naively one would expect that we have calculated the lighter of the two states but without including a larger basis of operators this cannot be shown for certain.

\begin{figure*}[!th]
\begin{minipage}[b]{0.49\linewidth}
    \includegraphics[width=0.8\hsize]{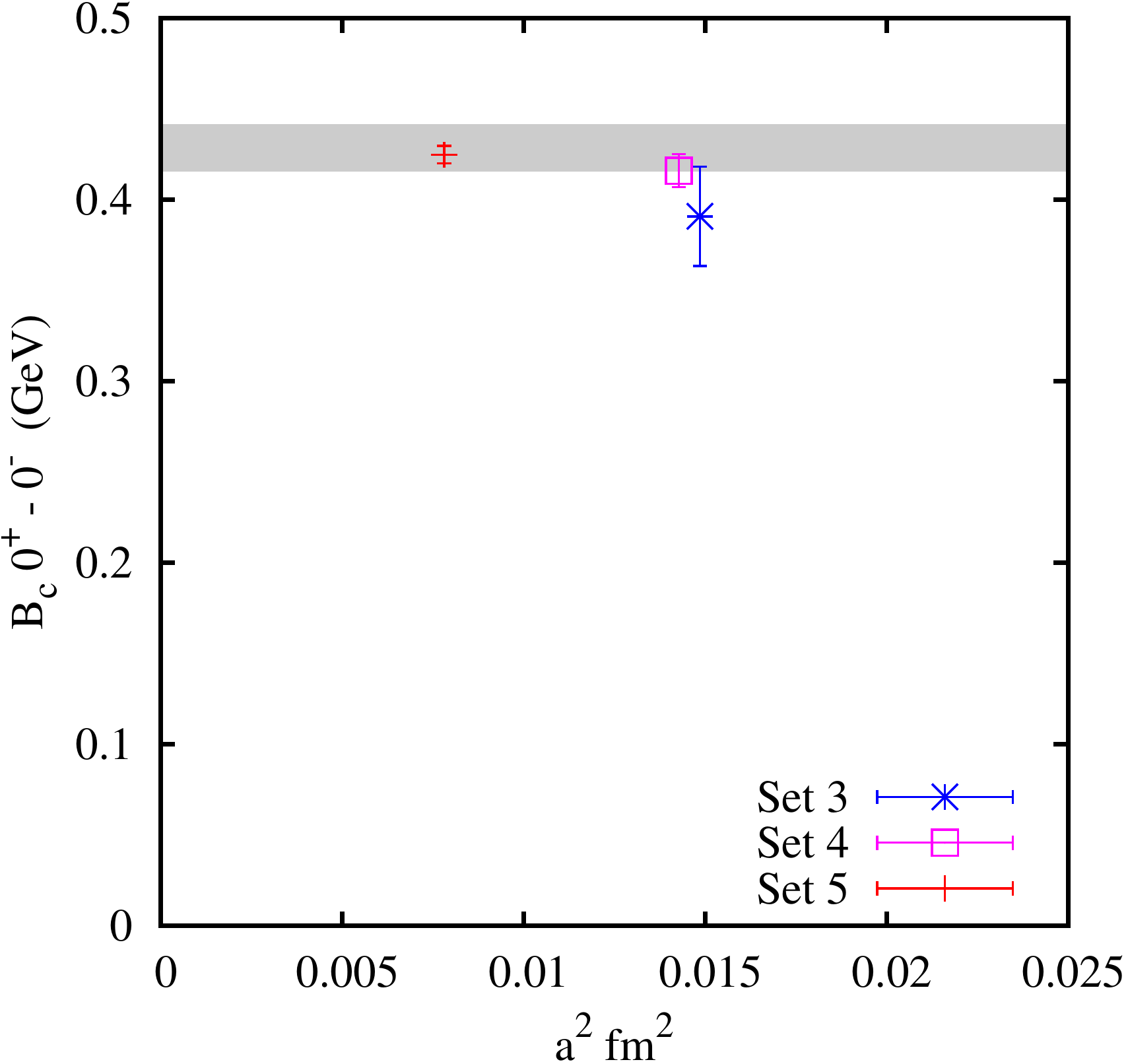}
    \caption{
    Fit and lattice data on the fine and coarse ensembles for the splitting $\Delta_{B_c}^{0^+-0^-}$. Errors include statistics and scale uncertainty only.
    }
    \label{fig:Bcscalar}
    %\end{figure}
\end{minipage}
\hspace{0.1cm}
\begin{minipage}[b]{0.49\linewidth}
    %\begin{figure}
    \includegraphics[width=0.8\hsize]{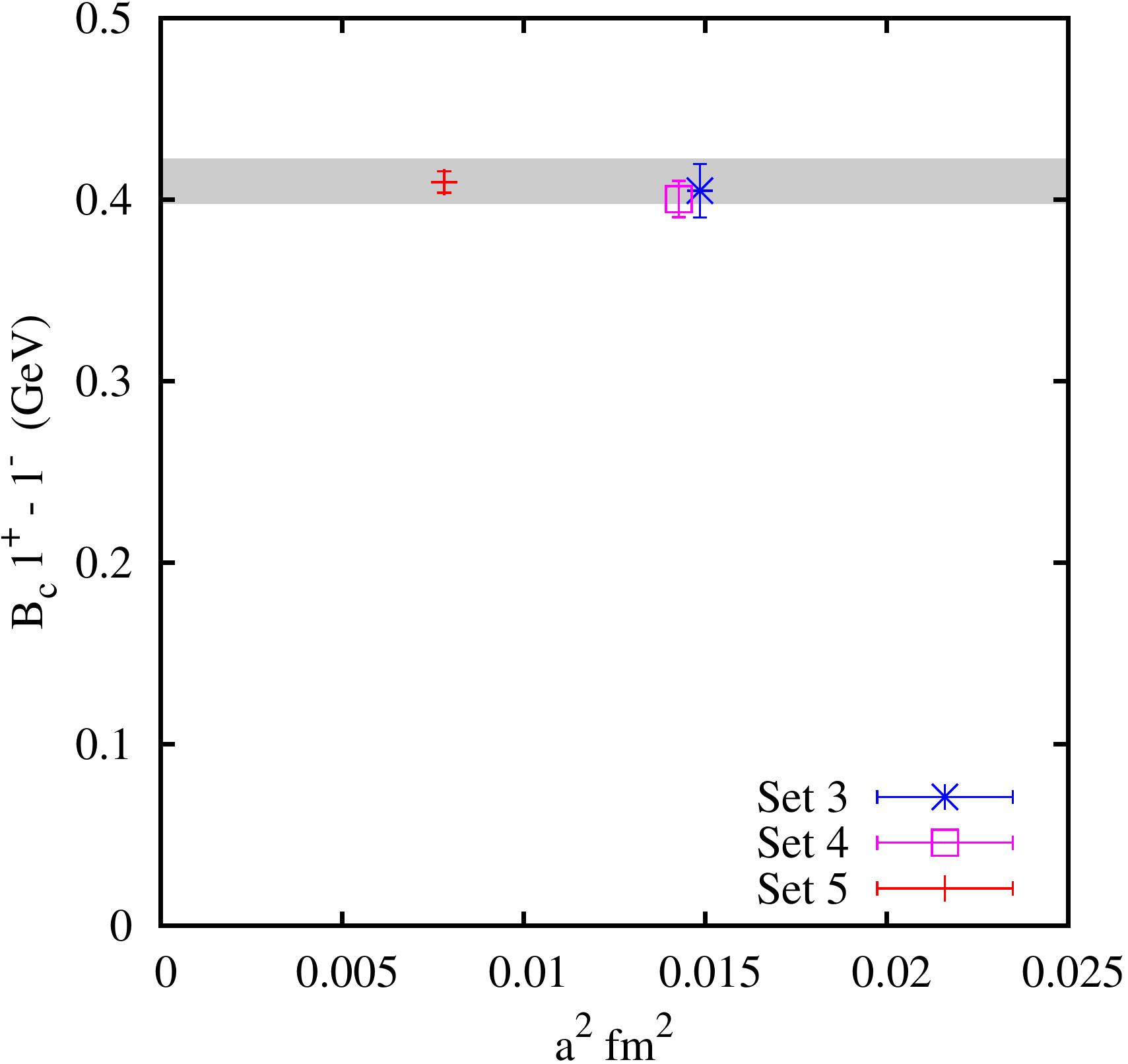}
    \caption{
    Fit and lattice data on the fine and coarse ensembles for the splitting $\Delta_{B_c}^{1^+-1^-}$. Errors include statistics and scale uncertainty only.
    }
    \label{fig:Bcaxial}
\end{minipage}
\end{figure*}

The results on coarse and fine ensembles 
for $\Delta_{B_c}^{0^+-0^-}$ and $\Delta_{B_c}^{1^+-1^-}$ are shown 
in Figs.~\ref{fig:Bcscalar} and~\ref{fig:Bcaxial} respectively. 
The results are fit using the same form as in Eq.~(\ref{eq:fitbchsextrap}) 
with a prior of 0.5(5) on the physical value and the same priors as before 
for other parameters. As in the case of the radially excited states, 
we estimate that errors from missing relativistic corrections to the 
NRQCD action 
will be 1 MeV and that other systematic errors will be negligible. 
Our results for the splittings are then: 
\begin{eqnarray}
\Delta_{B_c}^{0^+-0^-} &=& 429(13)(1) {\rm \ MeV}\\
\Delta_{B_c}^{1^+-1^-} &=& 410(13)(1) {\rm \ MeV},
\end{eqnarray}
where the first error is from the fit and the second is from NRQCD 
systematics.

%%%%%%%%%%%%%%%%%%%%%%%%%%%%%%%%%%%%%%%%%%%%%%%%%%%%%%%%%%%%%%%%%
%
\section{Discussion}
\label{sec:discussion}
\begin{widetext}
  \begin{figure*}
\begin{minipage}[b]{0.49\linewidth}
  \includegraphics[width=0.90\hsize]{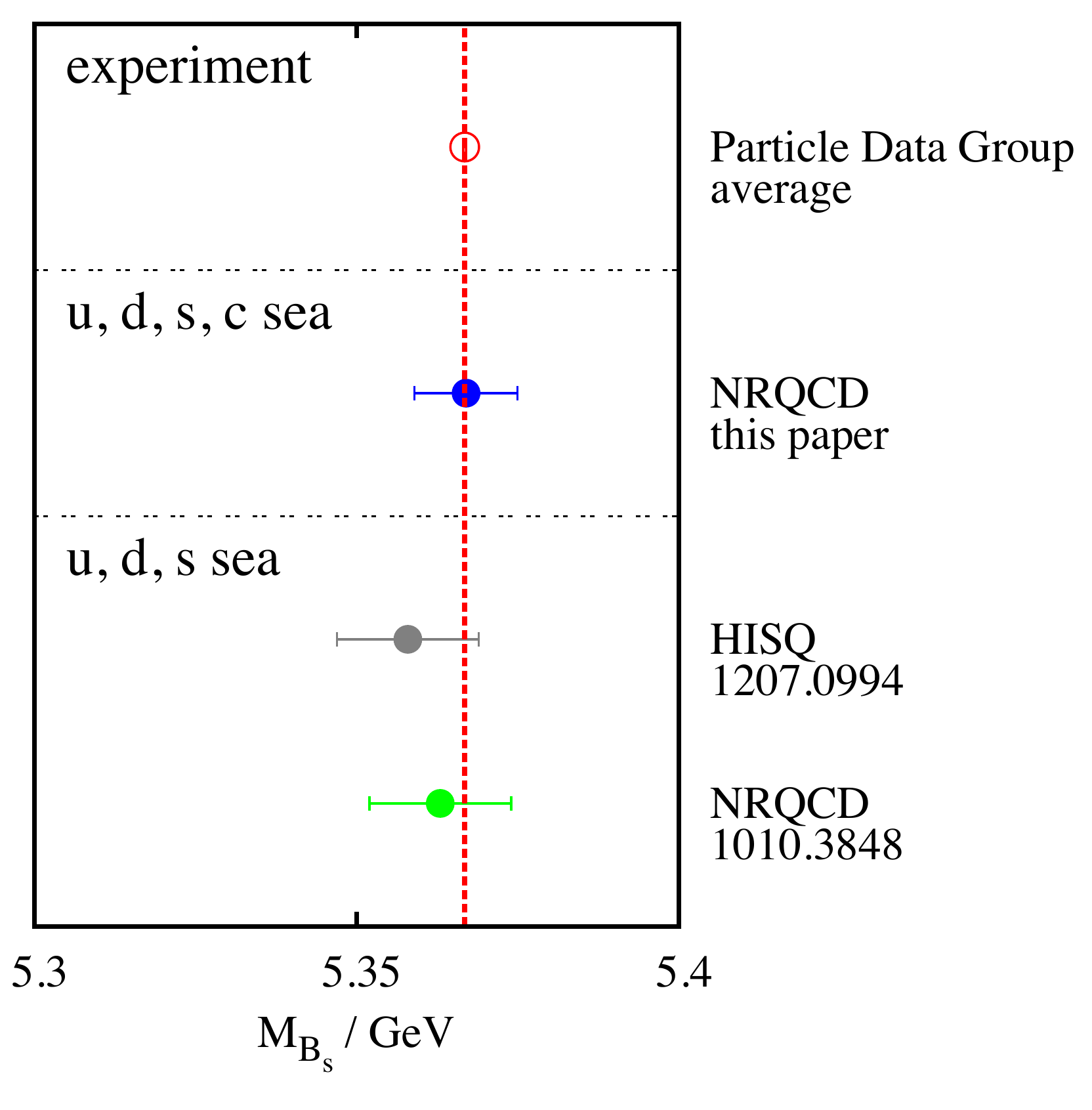}
  \caption{ A comparison of results for the $B_s$ meson 
  mass from different formalisms 
  for the $b$ quark in lattice QCD. 
  The experimental average value is given at the top 
  with accompanying vertical lines. 
  }
  \label{fig:mbsall}
  %\end{figure}
  %
\end{minipage}
\hspace{0.1cm}
\begin{minipage}[b]{0.49\linewidth}
  %\begin{figure}
  \includegraphics[width=0.90\hsize]{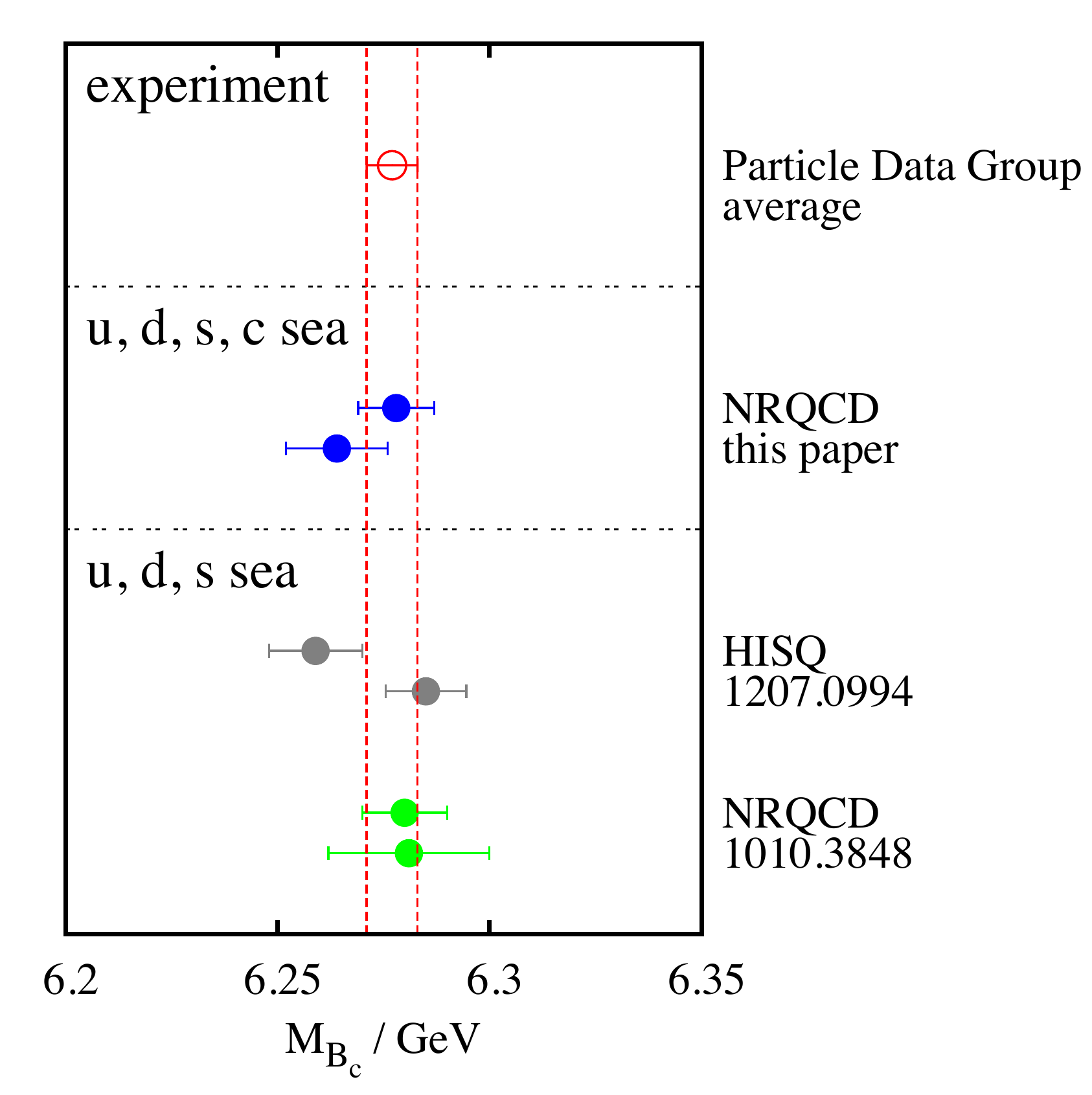}
  \caption{ A comparison of results for the $B_c$ meson 
  mass from different formalisms 
  for the $b$ quark in lattice QCD. 
  In each case the result from the hh method is 
  given above the result for the hs method. 
  The experimental average value is given at the top 
  with accompanying vertical lines. 
  }
  \label{fig:mbcall}
\end{minipage}
  \end{figure*}
\end{widetext}

\begin{comment}
    \begin{figure}
    \includegraphics[width=0.90\hsize]{./mbsall.pdf}
    \caption{ A comparison of results for the $B_s$ meson 
    mass from different formalisms 
    for the $b$ quark in lattice QCD. 
    The experimental average value is given at the top 
    with accompanying vertical lines. 
    }
    \label{fig:mbsall}
    \end{figure}
    %
    \begin{figure}
    \includegraphics[width=0.90\hsize]{./mbcall.pdf}
    \caption{ A comparison of results for the $B_c$ meson 
    mass from different formalisms 
    for the $b$ quark in lattice QCD. 
    In each case the result from the hh method is 
    given above the result for the hs method. 
    The experimental average value is given at the top 
    with accompanying vertical lines. 
    }
    \label{fig:mbcall}
    \end{figure}
\end{comment}

The results obtained here agree well with existing experiment and 
set improved levels of accuracy from a lattice QCD calculation. 

It is important to compare to other lattice QCD calculations as 
well as to experiment because different lattice QCD methods have 
different systematic errors, particularly if they use a different 
formalism for the quarks. Agreement then gives improved 
confidence in the error analysis.  In Figs.~\ref{fig:mbsall} 
and~\ref{fig:mbcall} we compare existing results for the masses 
of the $B_s$ and $B_c$ mesons from lattice QCD, in which 
the quark masses are fixed from bottomonium, the $\eta_c$ and 
the $\eta_s$. The comparison includes results from two 
very different formalisms for the $b$ quark: the NRQCD formalism 
used here and in~\cite{Gregory:2010gm} and the HISQ formalism 
in which an extrapolation up to the $b$ quark mass is made from 
lighter masses on lattices with a range of 
lattice spacings~\cite{McNeile:2011ng, McNeile:2012qf}.
The agreement between the different methods is good, within 
their total errors of around 10 MeV. 
%%%%%%%%%%%%%%%%%%%%%%%%%%%%%%%%%%%%%%%%%%%%%%%%%%%%%%%%%%%%%%%%%

%%%%%%%%%%%%%%%%%%%%%%%%%%%%%%%%%%%%%%%%%%%%%%%%%%%%%%%%%%%%%%%%%
%
\section{Conclusions}
\label{sec:conclusions}
%
%%%%%%%%%%%%%%%%%%%%%%%%%%%%%%%%%%%%%%%%%%%%%%%%%%%%%%%%%%%%%%%%%
\begin{figure}
\includegraphics[width=0.99\hsize]{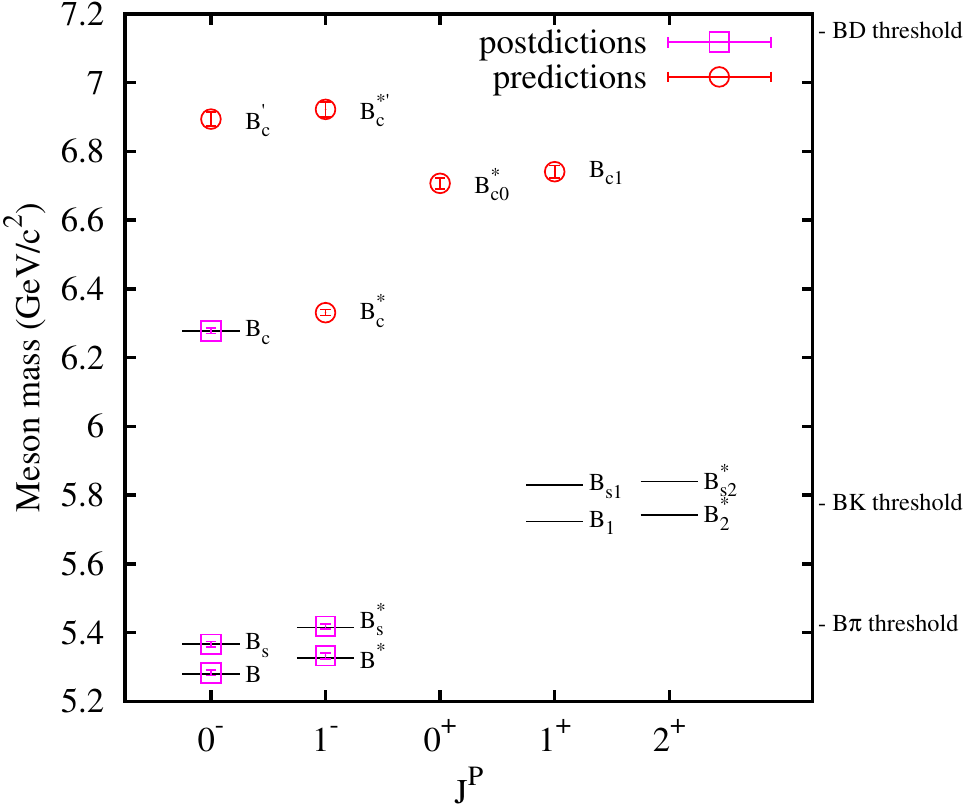}
\caption{The spectrum of gold-plated $B$, $B_s$ and $B_c$ 
meson states from this calculation, compared to experiment where 
results exist. Predictions are marked with open red circles. 
None of the meson masses included here were used to tune 
parameters of the action so all the masses are parameter-free 
results from lattice QCD. }
\label{fig:spectrum}
\end{figure}

We have presented results for the $B$ meson spectrum using a 
perturbatively improved NRQCD action, very high statistics and 
gluon field configurations with an improved gluon action and 
including 2+1+1 flavours of HISQ sea quarks. 
We have improved upon and extended the previous results in 
Ref.~\cite{Gregory:2010gm} and, combined with our study of 
the Upsilon spectrum in Ref~\cite{Dowdall:2011wh}, we have shown 
that our improved action gives accurate meson masses across a wide range of heavy mesons.
Where we can compare, we see no significant differences with the 
results of~\cite{Gregory:2010gm}, so that the inclusion of 
$c$ quarks in the sea has not produced any noticeable changes. 

The strongest improvement from our reduced systematic errors 
can be seen in the hyperfine splittings which were previously 
dominated by missing radiative corrections. 
Our errors are now 3-6 MeV, giving an even more stringent test 
against experiment than for the bottomonium hyperfine splitting.  
The high statistics used in our calculation (32k correlators with 3 quark smearings) 
allowed for the lightest $B$ states to be reliably extracted and, with 
the light sea quark masses now available, consistent results were 
obtained for a range of reasonable chiral fit functions. 
This demonstration is particularly important for future determinations 
of $f_B$ which are currently underway including ensembles with physical light quark masses.
The calculation showed that lattice QCD could successfully resolve the change in 
splitting between heavy-strange and heavy-light meson masses as 
the quark mass is increased from $c$ to $b$. 
The statistical precision of our correlators also allowed us to make the first QCD prediction of the radially excited $B_c$ states and two of the ``P-wave'' states.

An overview of our results for the B-meson spectrum is shown 
in Fig.~\ref{fig:spectrum} including the full error on each point. 
We find excellent agreement with the experimentally known pseudoscalar and vector states.
In summary our results are: 
$M_{B_s} - M_{B_l}=84(2)$ MeV, 
$M_{B_s}=5.366(8)$ GeV, 
$M_{B_c}=6.278(9)$ GeV, 
$M_{D_s}=1.9697(33)$ GeV, 
and $M_{D_s}-M_{D}=101(3)$ MeV.
Our results for the $B$ meson hyperfine splittings are 
$M_{B^*}-M_{B}=50(3)$ MeV 
and $M_{B_s^*}-M_{B_s}=52(3)$ MeV and 
we predict $M_{B_c^*}-M_{B_c}=54(3)$ MeV. 
Combining our results for the $B_c$ and the pseudoscalar radial 
splitting, we predict the mass of the $B_c^{'}$ to be 
$M_{B_c^{'}} = 6.894(19)\stat(8)\syst$ GeV. 
Combining the $B_c$, the hyperfine splitting and the vector radial splitting, we predict 
$M_{B_c^{*'}} = 6.922(19)\stat(8)\syst$ GeV. 
Our prediction for the 0$^+$ state is $M_{B_{c0}^{*}} = 6.707(14)\stat(8)\syst$ GeV.

\begin{figure*}
\includegraphics[width=0.7\hsize]{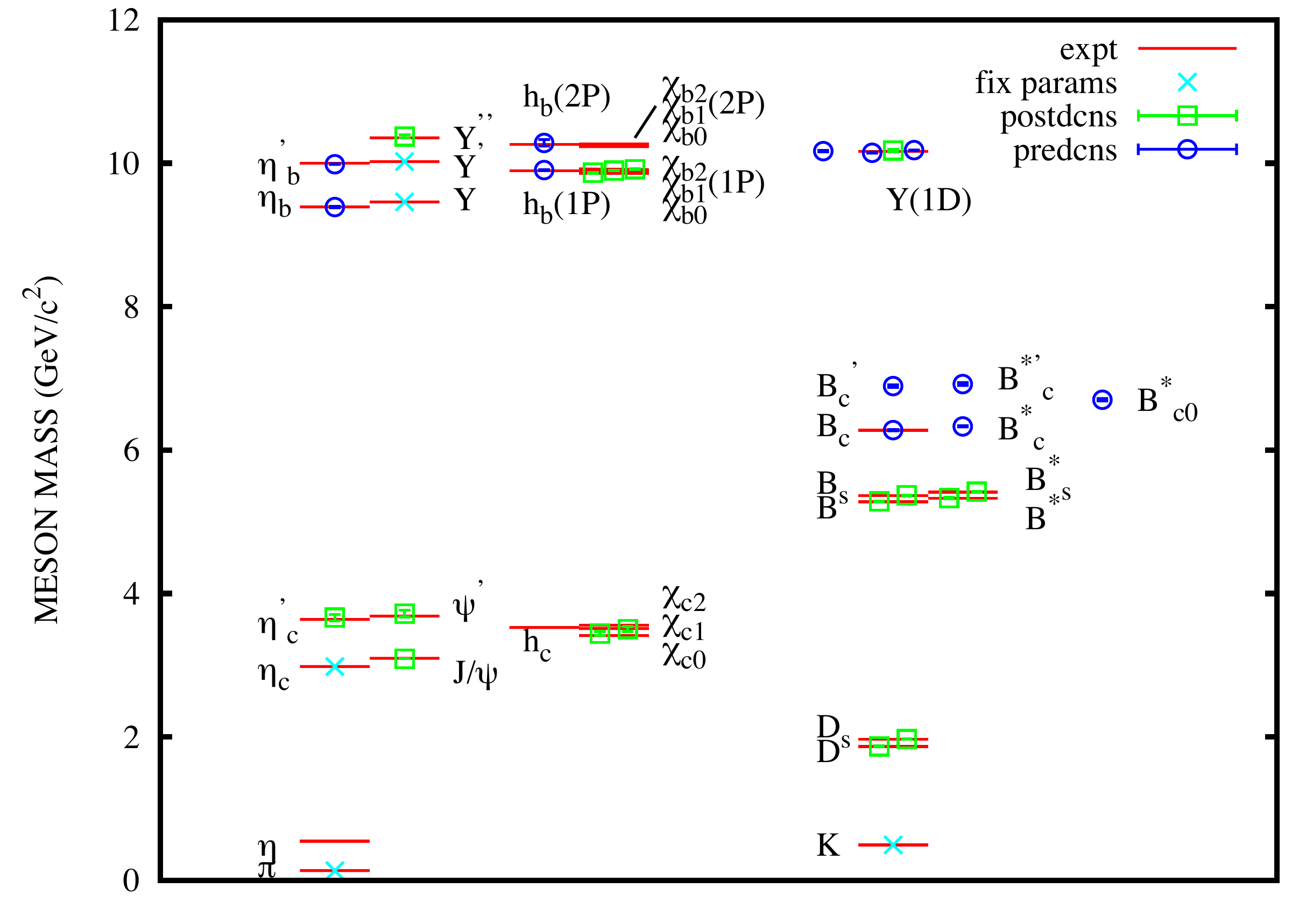}
\caption{The spectrum of gold-plated mesons comparing 
HPQCD lattice QCD results to experiment. 
We distinguish between results used to set the parameters 
of QCD, results obtained after experiment and results 
obtained before experimental values were available.  
}
\label{fig:goldspectrum}
\end{figure*}

Finally, in Fig.~\ref{fig:goldspectrum} we update 
the complete spectrum plot for gold-plated mesons 
to include the new results from this paper, as well 
as updated experimental values. 
This plot summarises the coverage and the predictive 
power of lattice QCD calculations. 

%%%%%%%%%%%%%%%%%%%%%%%%%%%%%%%%%%%%%%%%%%%%%%%%%%%%%%%%%%%%%%%%%
%
\section*{Acknowledgements}
\label{sec:acknowledgements}
%
%%%%%%%%%%%%%%%%%%%%%%%%%%%%%%%%%%%%%%%%%%%%%%%%%%%%%%%%%%%%%%%%%+
We are grateful to the MILC collaboration for the use of their 
gauge configurations and to J. Laiho, P. Lepage and C. Monahan for useful discussions. 
The results described here were obtained using the Darwin Supercomputer 
of the University of Cambridge High Performance 
Computing Service as part of the DiRAC facility jointly
funded by STFC, the Large Facilities Capital Fund of BIS 
and the Universities of Cambridge and Glasgow. 
This work was funded by STFC. 

%\appendix

%\bibliographystyle{ieeetr}
\bibliography{hl_bib}

\end{document}